\documentclass{aa}  %

\usepackage{graphicx}
\usepackage{txfonts}
\begin{document}

   \title{A study of carbon-rich post-AGB stars in the
   Milky Way to understand the production of carbonaceous dust 
   from evolved stars}


   \author{S. Tosi\inst{1,2}, D. Kamath\inst{2,3,4}, F. Dell'Agli\inst{2}, H. Van Winckel\inst{5},
           P. Ventura\inst{2,6}, T. Marchetti\inst{7}, E. Marini\inst{2}, M. Tailo\inst{8}}

   \institute{Dipartimento di Matematica e Fisica, Università degli Studi Roma Tre, 
              via della Vasca Navale 84, 00100, Roma, Italy \and
              INAF, Observatory of Rome, Via Frascati 33, 00077 Monte Porzio Catone (RM), Italy \and
              School of Mathematical and Physical Sciences, Macquarie University, Balaclava Road, Sydney, NSW 2109, Australia \and
              Research Centre for Astronomy, Astrophysics and Astrophotonics, Macquarie University, Balaclava Road, Sydney, NSW 2109, Australia  \and
              Institute of Astronomy, K.U.Leuven, Celestijnenlaan 200D bus 2401, B-3001 Leuven, Belgium \and
              Istituto Nazionale di Fisica Nucleare, section of Perugia, Via A. Pascoli snc, 06123 Perugia, Italy \and
              European Southern Observatory, Karl-Schwarzschild-Strasse 2, 85748 Garching bei München, Germany  \and
              Dipartimento di Fisica e Astronomia Augusto Righi, Università degli Studi di Bologna, Via Gobetti 93/2, I-40129 Bologna, Italy
              }

   \date{Received September 15, 1996; accepted March 16, 1997}


 \abstract
   {Knowledge of the $\it{GAIA}$\, DR3 parallaxes of Galactic post-asymptotic giant branch (AGB) stars 
    makes it possible to exploit these objects as tracers of AGB evolution, nucleosynthesis, and dust production as well as  to use them to shed new light on still poorly known 
    physical processes experienced by AGB stars.}
   {The goal of this study is to reconstruct the evolution and the dust formation processes during the final AGB phases of a sample of carbon-rich, post-AGB Galactic stars, with particular attention to the determination of the past mass-loss history.}
   {We study the IR excess of Galactic sources classified as post-AGB single stars by means of dust formation modelling where dust grains form and grow in a static wind
and expand from the surface of the star. The method is applied to various evolutionary stages of the final AGB phase of stars with different masses and metallicities. The results from a spectral energy distribution (SED) fitting are used to infer information on mass loss, efficiency of dust formation, and wind dynamics.}
   { The detailed analysis of the SED of the sources investigated, which included the
   derivation of the luminosities and the dust properties, allows us
   to confirm previous results, mostly based on the surface chemical composition, that
   most of the investigated sources  descend from low-mass (${\rm M}<1.5~{\rm M}_{\odot}$) progenitors that reached the C-star stage. Metal-poor carbon stars are characterised by higher IR excesses with respect to their more metal-rich
    counterparts of similar luminosity due to a higher surface carbon-to-oxygen excess. This work confirms previous conclusions based on a limited sample of
    carbon-rich post-AGB objects in the Magellanic Clouds, namely that more luminous stars descending from higher-mass progenitors are generally more opaque due to shorter evolutionary timescales that place the dust shell closer to the central object. 
    Through the study of the dynamics of the outflow and results from stellar evolution modelling, we find that the mass-loss rate at the tip of the AGB phase of metal-rich low-mass carbon stars is approximately $1-1.5\times 10^{-5}~{\rm M}_{\odot}/$yr, whereas in the metal-poor domain $\dot{\rm M } \sim 4-5\times 10^{-5}~{\rm M}_{\odot}/$yr is required.  These results indicate the need for an upwards revision of the
    theoretical mass-loss rates of low-mass carbon stars in the available literature,
    which in turn require a revised determination of carbon dust yields by AGB stars.}
   {}


   \titlerunning{Evolution and dust formation in carbon stars of the Galaxy}
   \authorrunning{Tosi et al.}
   \maketitle
%

\section{Introduction}
All stars with initial masses in the range of $1-8~\rm{M}_{\odot}$ pass through the asymptotic giant branch (AGB) phase. The AGB stars eject large amounts of reprocessed gas into the interstellar medium, which is why they are regarded as one of the most important contributors to the interstellar medium, playing a crucial role in the evolution of the abundances of chemical species in the Milky Way \citep{romano10, kobayashi20} and in Local Group galaxies \citep{vincenzo16}.  The AGB stars are also efficient dust producers
due to the thermodynamic conditions of their winds, which prove to be favourable
environments for the condensation of gaseous molecules into solid particles. 
Notably, the stars have provided a significant contribution to the overall
dust budget in the Magellanic Clouds \citep[MC;][]{boyer12, matsuura11, matsuura13}.

The AGB stage is characterised by two main physical phenomena that modify the surface chemistry of a star: The third dredge-up (TDU; Iben 1974) and hot bottom burning (HBB; \citet{hbb}). The first event favours a gradual enrichment in the surface $^{12}$C and $s$-process elements, while HBB leads to a change in the surface chemical composition that affects
the equilibrium abundances of the different chemical species, such as C, 
N, and Li, corresponding to the temperature at the base of the envelope. The structural and evolutionary properties of AGB stars are still not fully understood, owing to the poor knowledge of some physical mechanisms, mainly convection and mass loss, that play a crucial role in the evolution of these stars \citep{ventura05a, ventura05b}. 

The study and characterisation of post-AGB stars has considerably improved the general understanding of the evolution of AGB stars. Indeed, the chemical composition of the latter objects represents the final outcome of the AGB chemical evolution and the associated internal enrichment process, which provides
valuable information on the relative efficiency of TDU and HBB to determine
the final chemical composition. On the observational side, optical spectra 
of post-AGBs are dominated by atomic transitions and provide a unique opportunity for deriving accurate photospheric chemical abundances of a wide range of species, such as CNO, $\alpha$, Fe-peak, and $s$-process elements 
\citep[][and references therein]{devika2020, devika22c}. Furthermore, thanks to the peculiar shape of the spectral energy distribution (SED) of post-AGB stars, which shows a typical double-peaked behaviour \citep{hans03,devika14, devika15}, it is possible to disentangle the contribution of the dust shell from the emission of the central object. 
Therefore, the study of the IR excess of post-AGB stars offers a valuable opportunity for deducing the properties of the circumstellar dust in terms of mineralogy and optical thickness as well as the current distance of the dusty shell from the central stars.  Indeed, the observations of post-AGB stars can be interpreted on the basis of dust formation
modelling, which some research teams have recently incorporated into the treatment of the AGB
phase. In particular, the investigations by \citet{nanni13, nanni14, ventura12, ventura14}
have provided a description of the dust formation process in the winds of AGB stars with
different masses and metallicities by considering a static wind expanding from the 
surface of the star and then entering the dust condensation zone where dust grains
form with a growth rate primarily determined by the density of the wind and the
surface chemical composition of the stars.
Comparison between the observation of post-AGB stars with results from dust formation modelling allows for the reconstruction of the mass-loss and dust production mechanisms during the late AGB phases, which generally prove to be the most relevant for assessing the dust yields expected from AGB stars. 

Our recent studies have been aimed at investigating and interpreting the observations of post-AGB stars in order to provide a characterisation of these objects and to reconstruct the evolution and the dust formation process in the wind of AGB stars, particularly during the late AGB stages. 
 We selected sources classified as single post-AGB objects based on the analysis
of the evolution of their radial velocities.
Tosi et al. (2022; hereafter T22) investigated a sub-sample of single post-AGB stars in the Magellanic Clouds listed in \cite{devika14} and \cite{devika15}. These studies presented the SEDs of single post-AGB stars in the MC and pointed to the shell-type structure of their dust distribution, thus suggesting their single evolutionary nature. The analysis proposed in T22 allowed for the characterisation of the dust responsible for the IR excess observed in the SED of the individual sources. Using this information, the authors of T22 were able to reconstruct the transition from the late AGB phases to the present time and to describe the variation of the mass-loss rate experienced by the stars to derive the properties of the outflow. The methodology proposed in T22 could be applied to a small sample made up of 13 objects. 

Kamath et al. (2022, hereafter K22) recently studied a sample of 31 single post-AGB stars in the Milky Way for which parallaxes are available as part of the third data release (DR3) of the ESA Satellite {\it Gaia} \citep[][]{gaia1,lindegren2021}. More recently, Kamath et al. (2023, hereafter K23), on the basis of the effective temperatures, luminosities, and surface chemical composition, used results from stellar evolution modelling to characterise the individual sources presented in K22 to reconstruct their past evolutionary history and to deduce the mass and the epoch when their progenitors formed. The study  in K22 allowed us to extend the analysis applied to MC sources in T22 to the Galactic counterparts, which had previously been inhibited by poor knowledge of their distances. 

\citet{flavia22} recently focused on the oxygen-rich stars in the sample from K22, which show no trace of  an $s$-process nor carbon enrichment. The study by \citet{flavia22}, based on the observed IR excess of the individual sources, allowed us to reconstruct the late AGB stages of low-mass M-type stars and of the higher-mass counterparts that experienced HBB in order to investigate the role of radiation pressure on the wind dynamics of oxygen-rich AGB and post-AGB stars and to derive an estimate of the mass-loss rate across the final AGB stages through the contraction of the central
star to the post-AGB phase.

In this paper, we consider the much wider sample of carbon-rich single post-AGB stars that also show signatures of $s$-process enhancement (see K22 and K23, and references therein). The C-star nature of these objects is further supported by the signature of the $21~\mu$m and $30~\mu$m features in their SEDs \citep{kwok13}. The $21~\mu$m feature remains unidentified, although it appears to be associated with the presence of complex hydrocarbons (e.g. Volk et al. 2020), while the $30~\mu$m feature most likely arises from MgS grains (Goebel \& Moseley 1985; Sloan et al. 2014). 
We aim to interpret the observed IR excess in order to investigate the timing of the AGB to post-AGB phase transition of carbon stars, the mass-loss rates experienced near the end of the AGB phase and during the contraction of the central star to the post-AGB, and the dynamics of the outflow from the time when the dust responsible for the currently observed IR emission was formed until the present. The C-star sample from K22 is particularly rich and enables the study of the role of mass and metallicity on the AGB and post-AGB evolution and on the dust formation process.

The paper is structured as follows: In Section \ref{method}, we describe the methodology followed to characterise the individual sources examined. The results obtained from SED fitting via the comparison of the observations with synthetic modelling are listed in Section \ref{sedfit}. The discussion of the role of mass and chemical composition on the properties of the outflow of post-AGB stars and of the dust formation process is given in Section \ref{disc}. In Section \ref{id817}, we examine the evolution and dynamics of two sources whose observed IR excess bracket the observed IR emission of the majority of the C-stars in the sample. Finally, the conclusions are given in Section \ref{concl}.

\begin{figure*}
\begin{minipage}{0.32\textwidth}
\resizebox{1.\hsize}{!}{\includegraphics{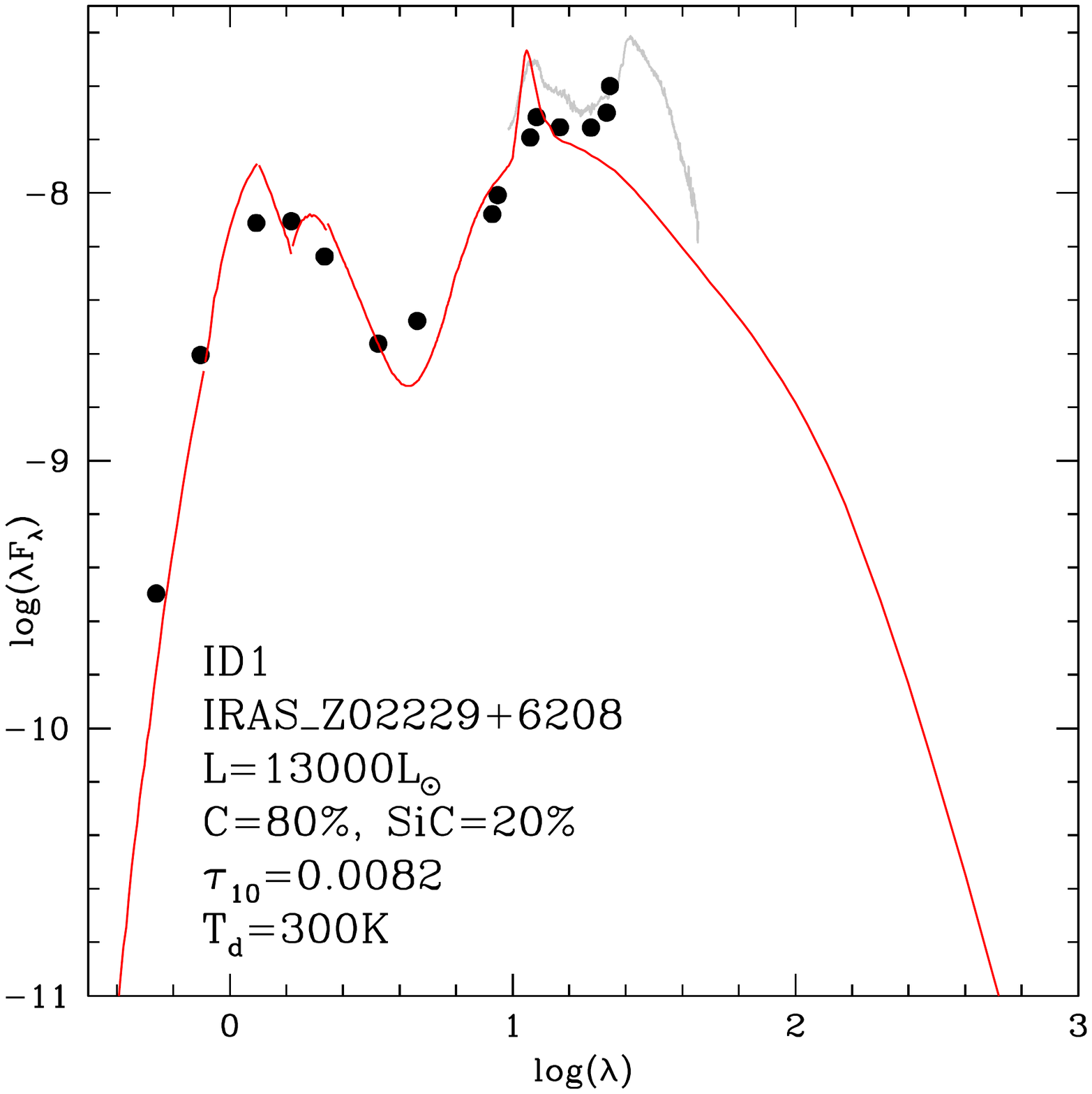}}
\end{minipage}
\begin{minipage}{0.32\textwidth}
\resizebox{1.\hsize}{!}{\includegraphics{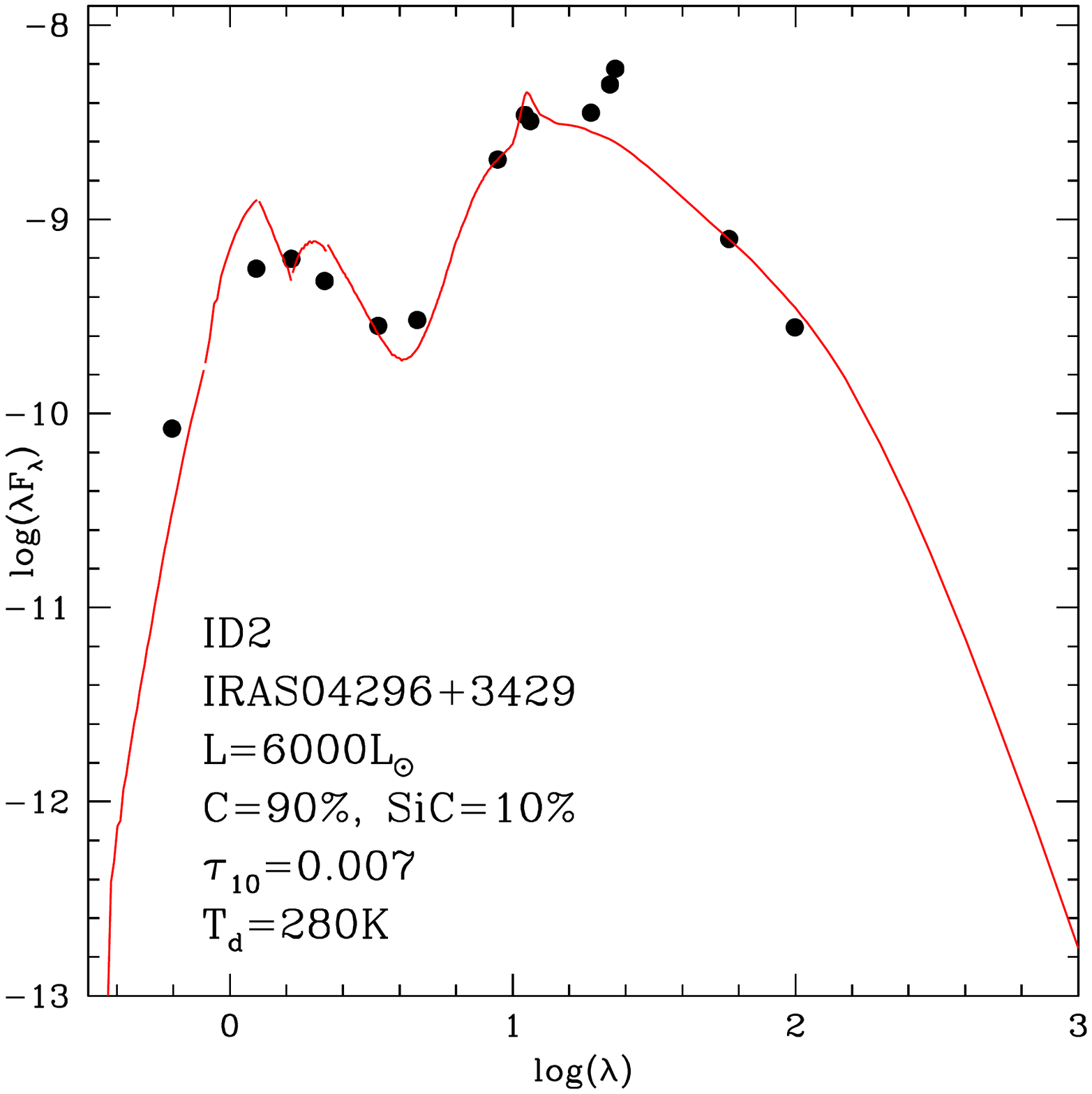}}
\end{minipage}
\begin{minipage}{0.32\textwidth}
\resizebox{1.\hsize}{!}{\includegraphics{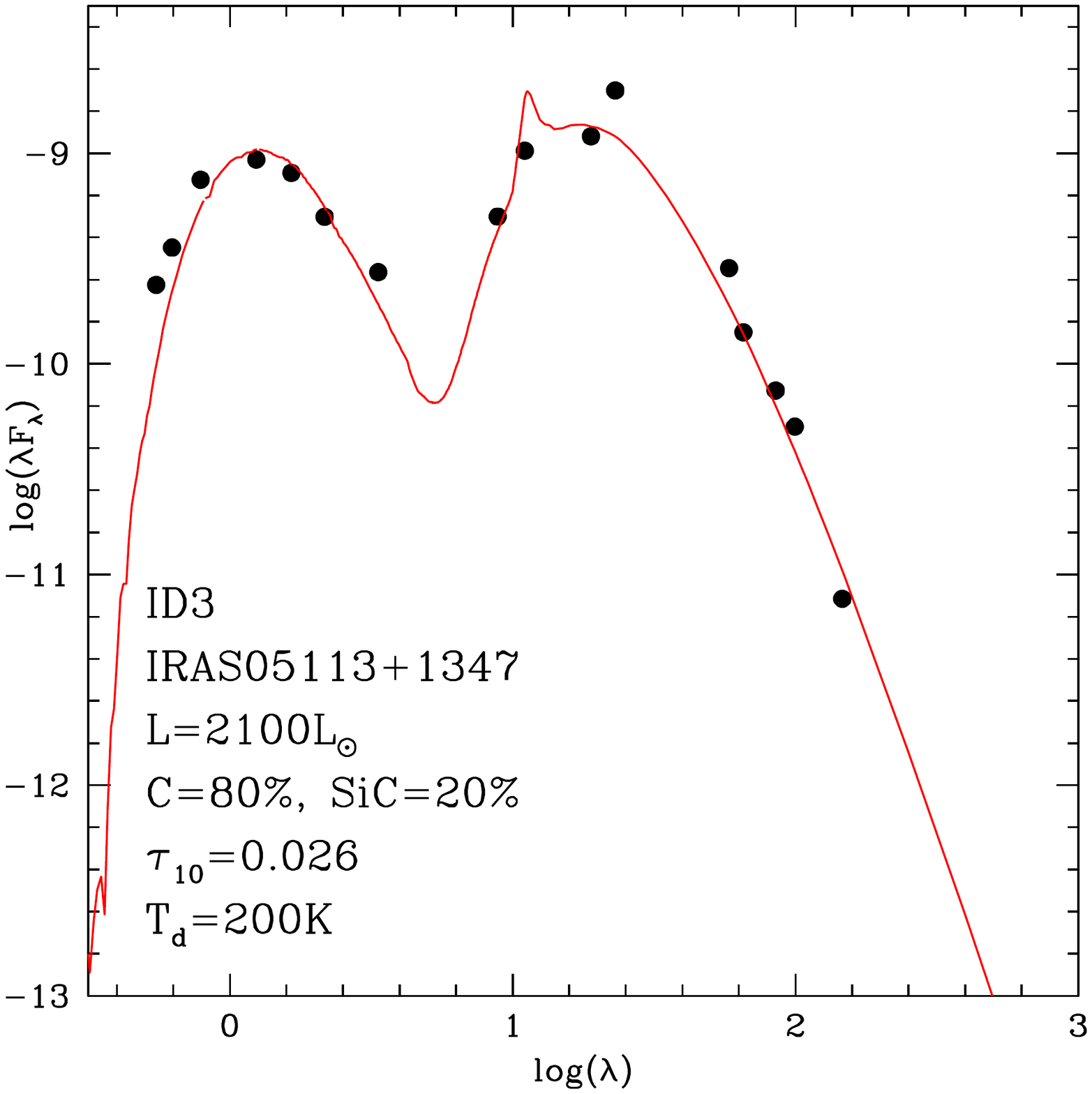}}
\end{minipage}
\vskip-60pt
\begin{minipage}{0.32\textwidth}
\resizebox{1.\hsize}{!}{\includegraphics{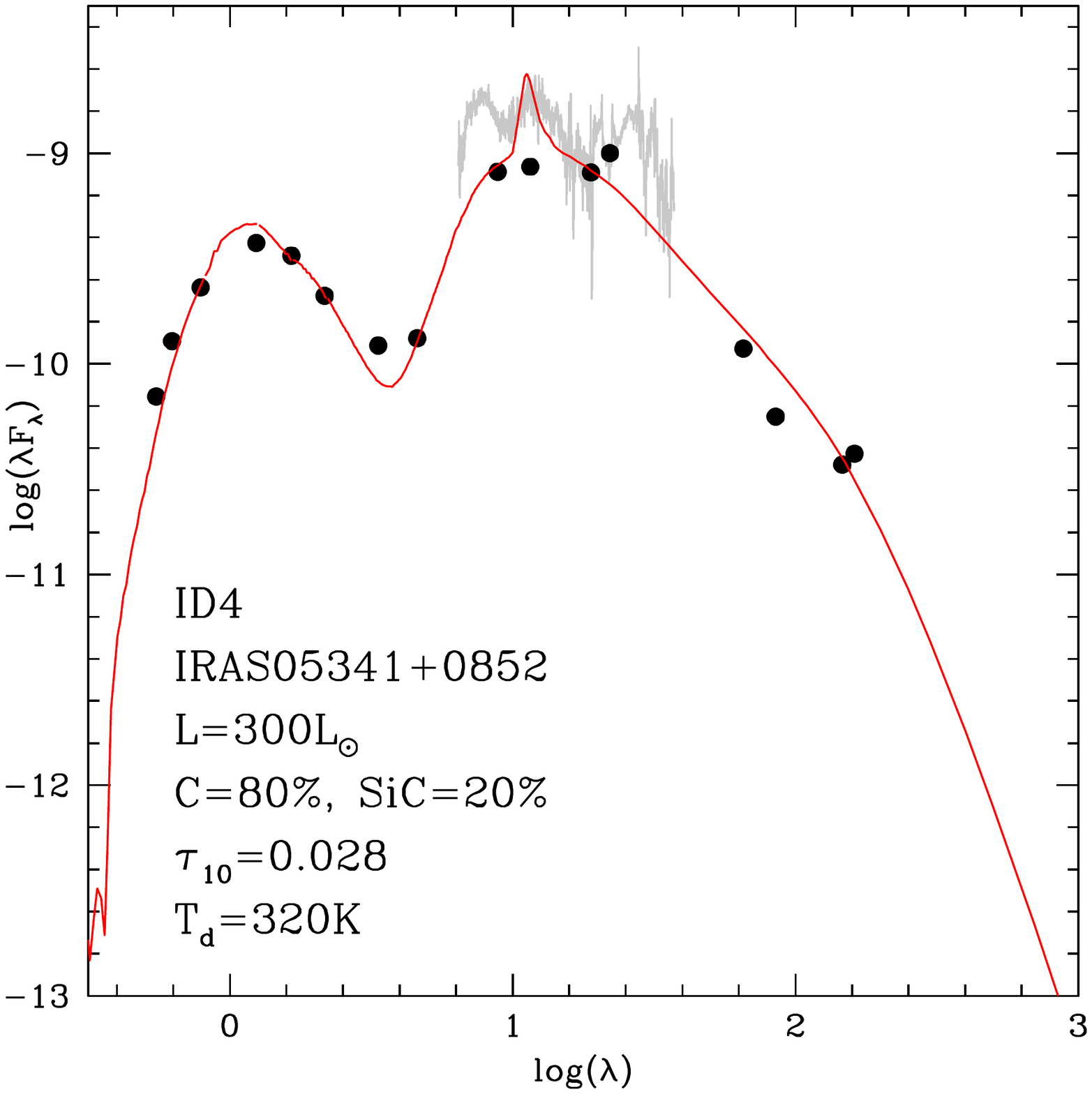}}
\end{minipage}
\begin{minipage}{0.32\textwidth}
\resizebox{1.\hsize}{!}{\includegraphics{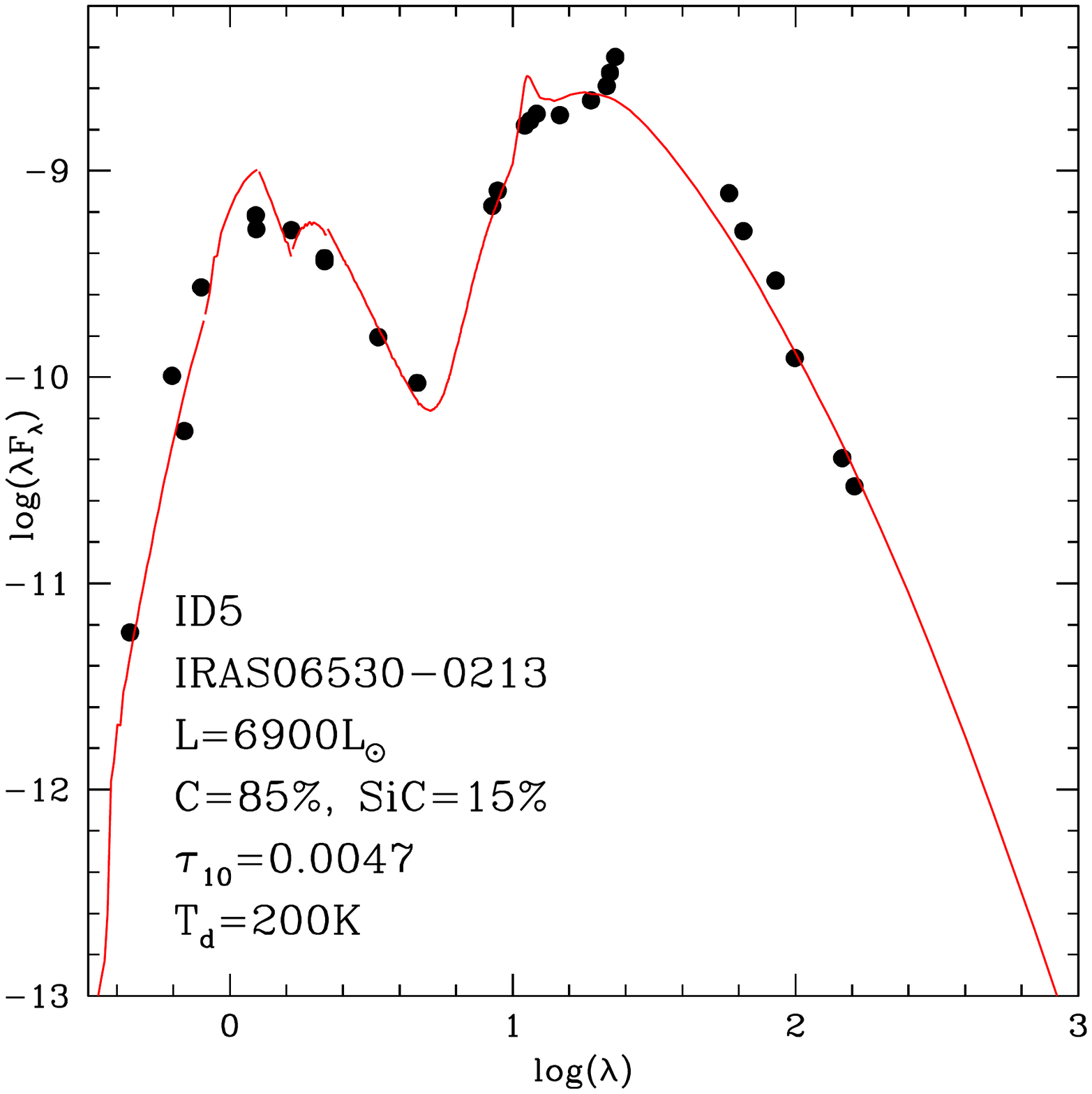}}
\end{minipage}
\begin{minipage}{0.32\textwidth}
\resizebox{1.\hsize}{!}{\includegraphics{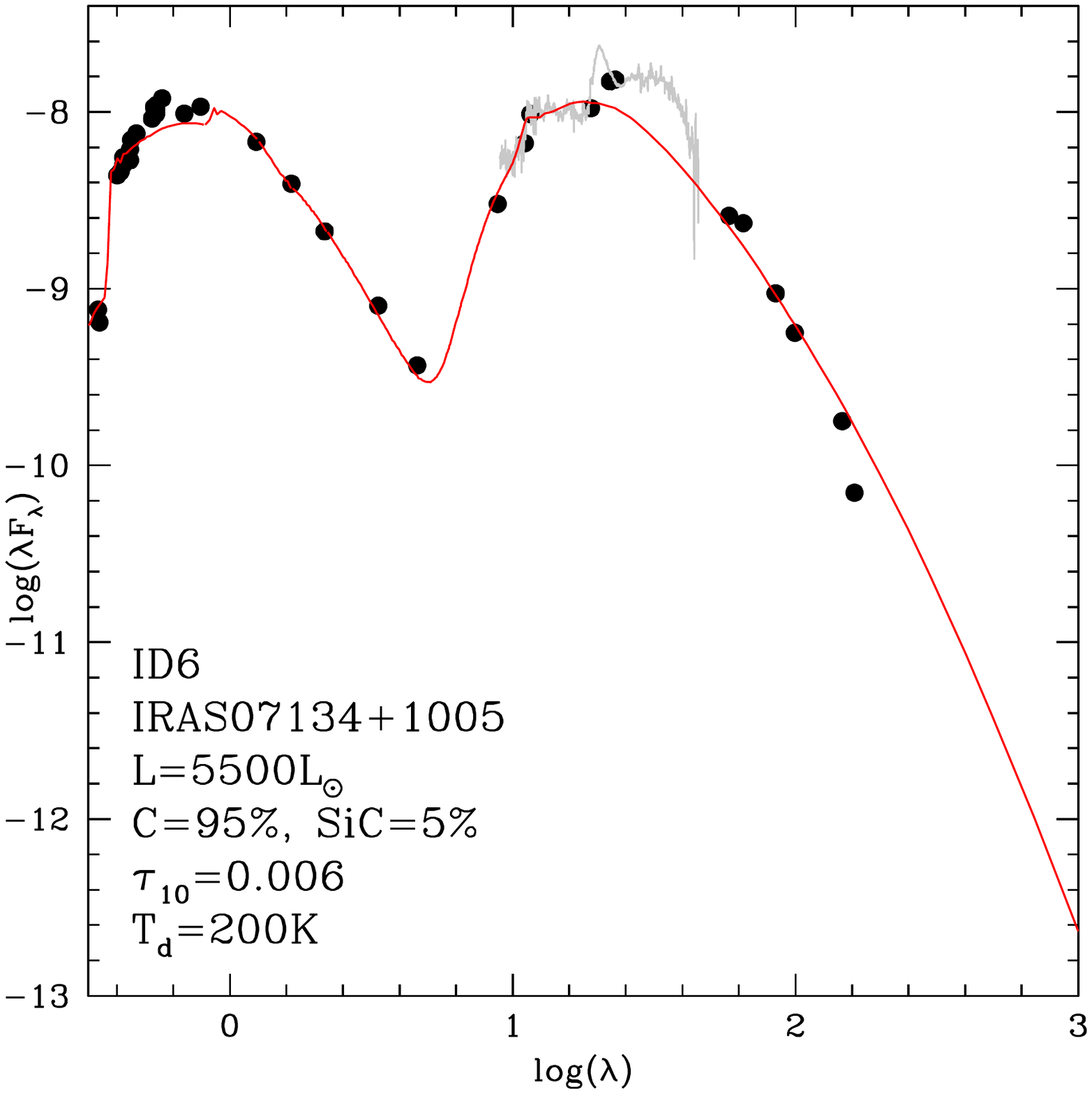}}
\end{minipage}
\vskip-60pt
\begin{minipage}{0.32\textwidth}
\resizebox{1.\hsize}{!}{\includegraphics{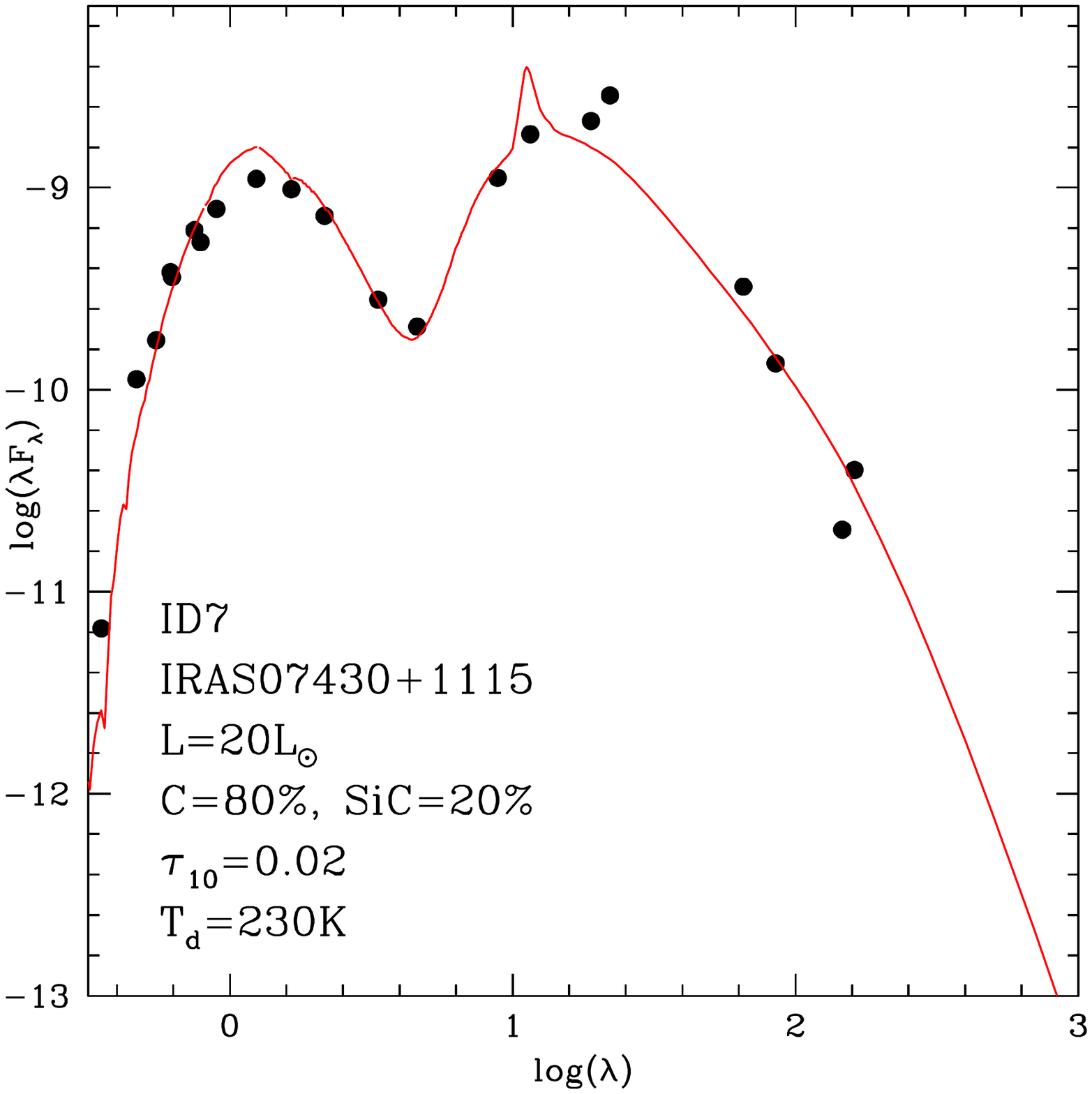}}
\end{minipage}
\begin{minipage}{0.32\textwidth}
\resizebox{1.\hsize}{!}{\includegraphics{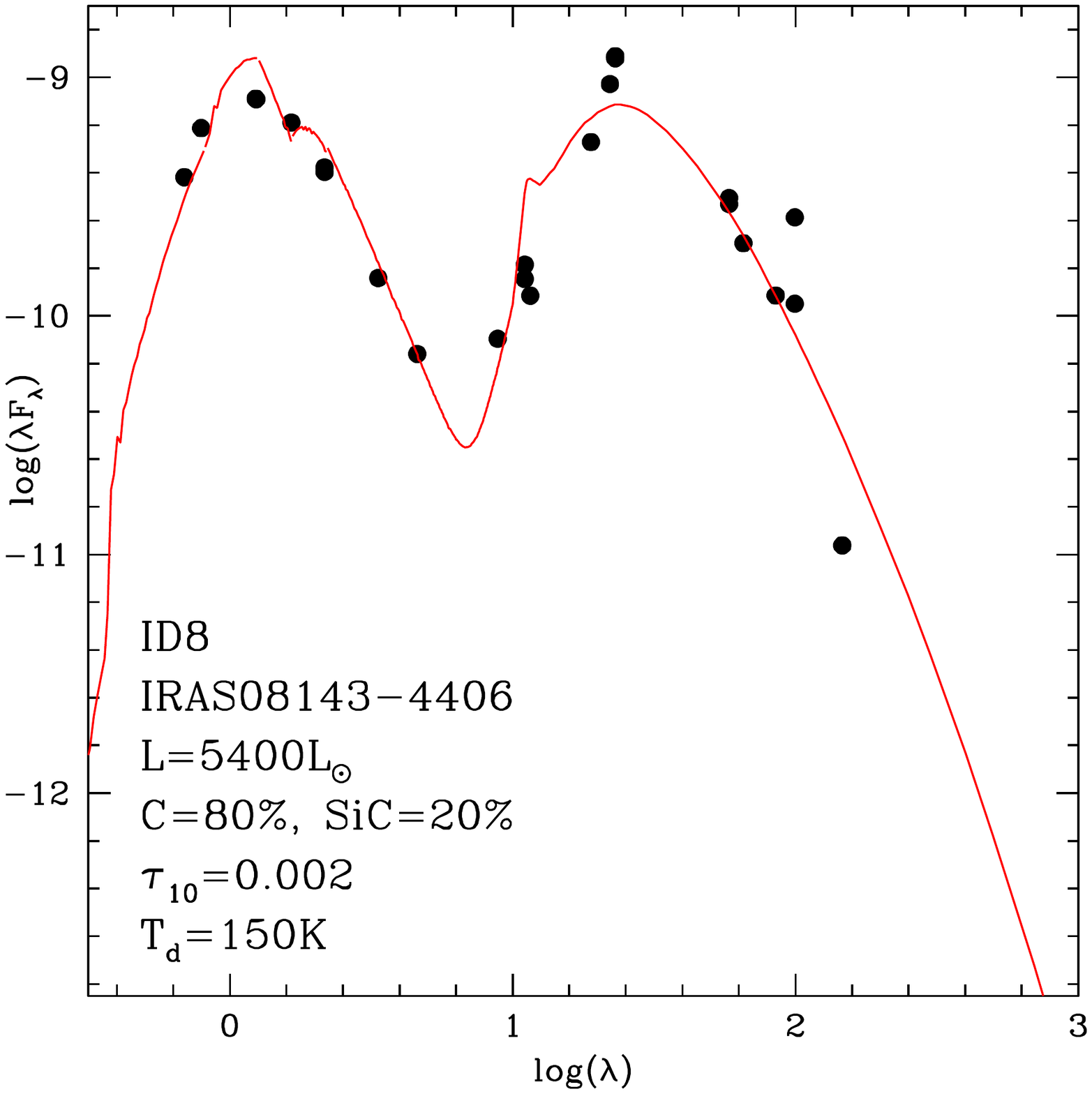}}
\end{minipage}
\begin{minipage}{0.32\textwidth}
\resizebox{1.\hsize}{!}{\includegraphics{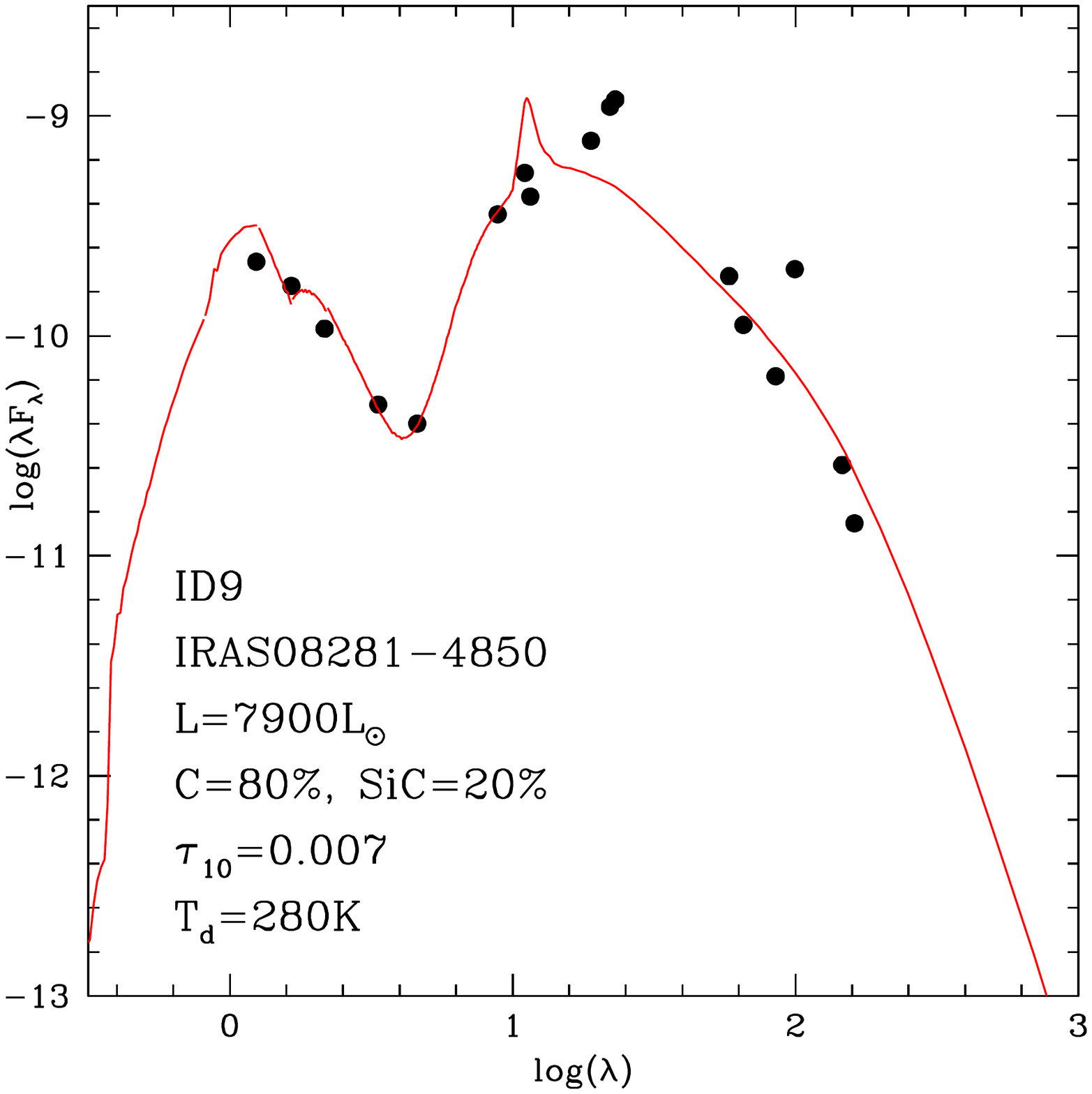}}
\end{minipage}
\vskip-40pt
\caption{Optical and IR data (black points) of Galactic sources classified as single C-rich post-AGB stars presented in \citet{devika22}, which we interpret as surrounded by  carbonaceous dust in this study. The red lines indicate the best-fit model obtained using the DUSTY code (see Section~\ref{sedfit}). The grey line shows the SWS spectra from \citet{sloan03}, when available. The derived stellar and dust parameters from this study for each source are shown in the different panels.}
\label{fsed}
\end{figure*}

\begin{figure*}
\begin{minipage}{0.32\textwidth}
\resizebox{1.\hsize}{!}{\includegraphics{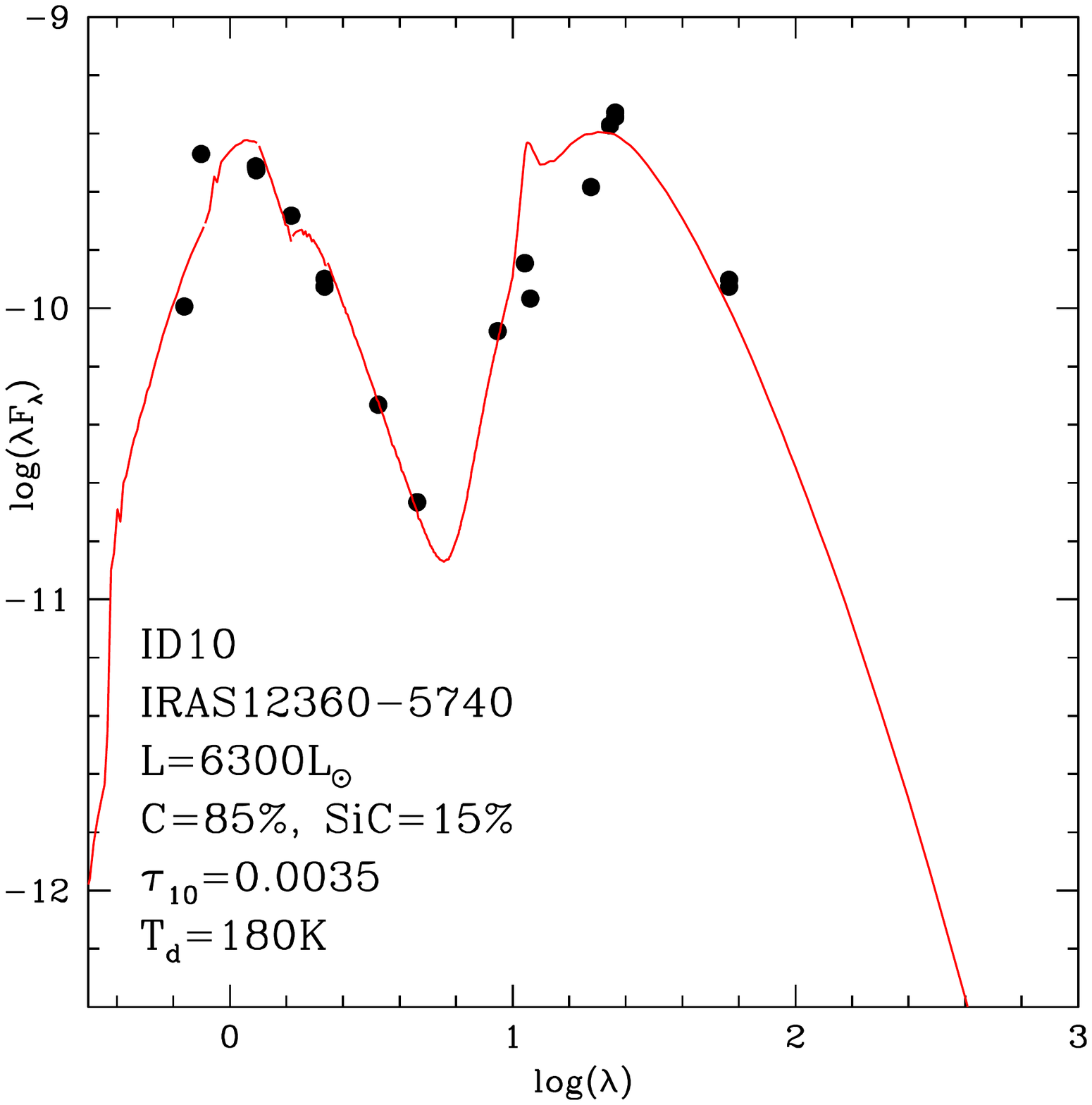}}
\end{minipage}
\begin{minipage}{0.32\textwidth}
\resizebox{1.\hsize}{!}{\includegraphics{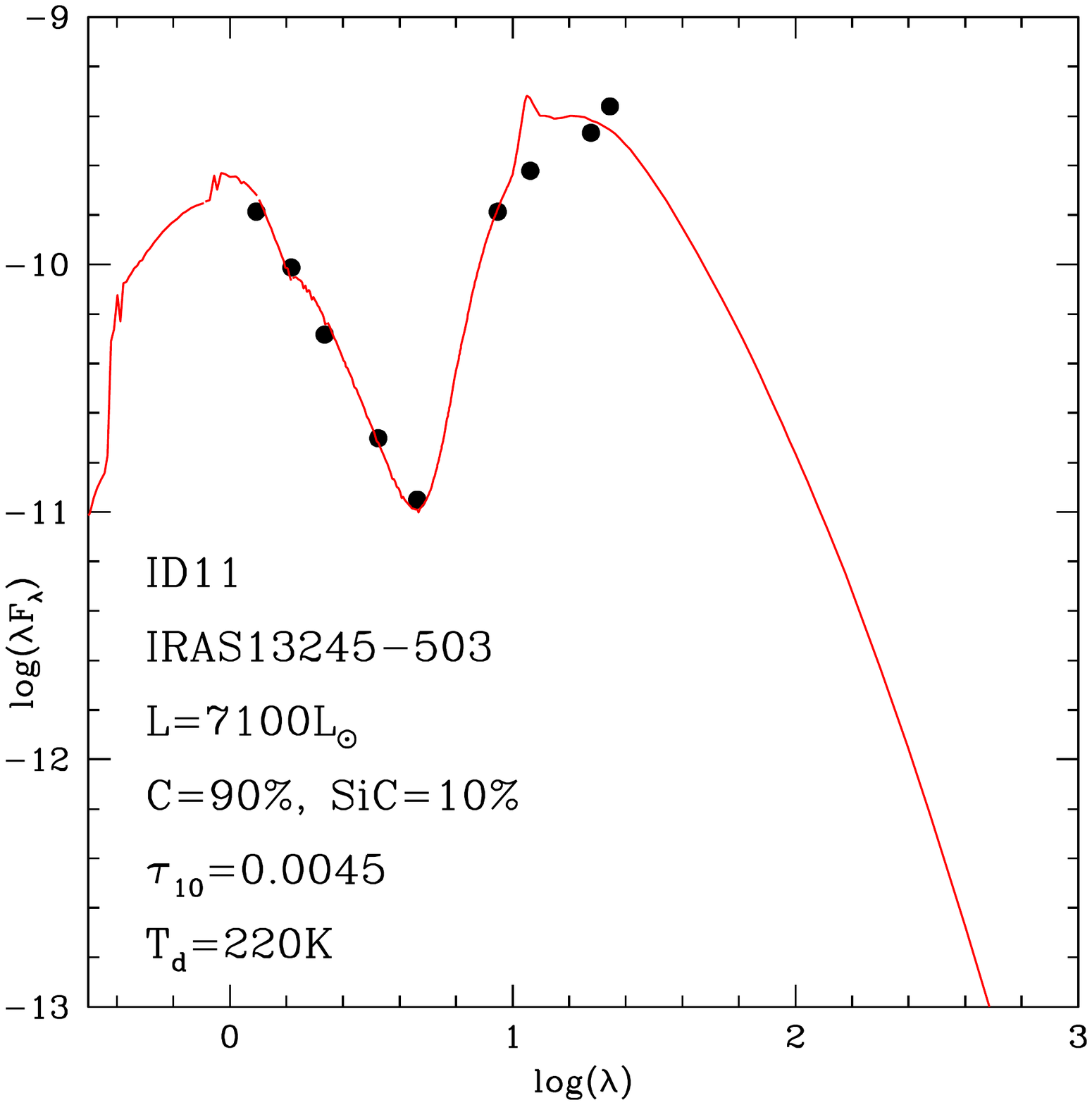}}
\end{minipage}
\begin{minipage}{0.32\textwidth}
\resizebox{1.\hsize}{!}{\includegraphics{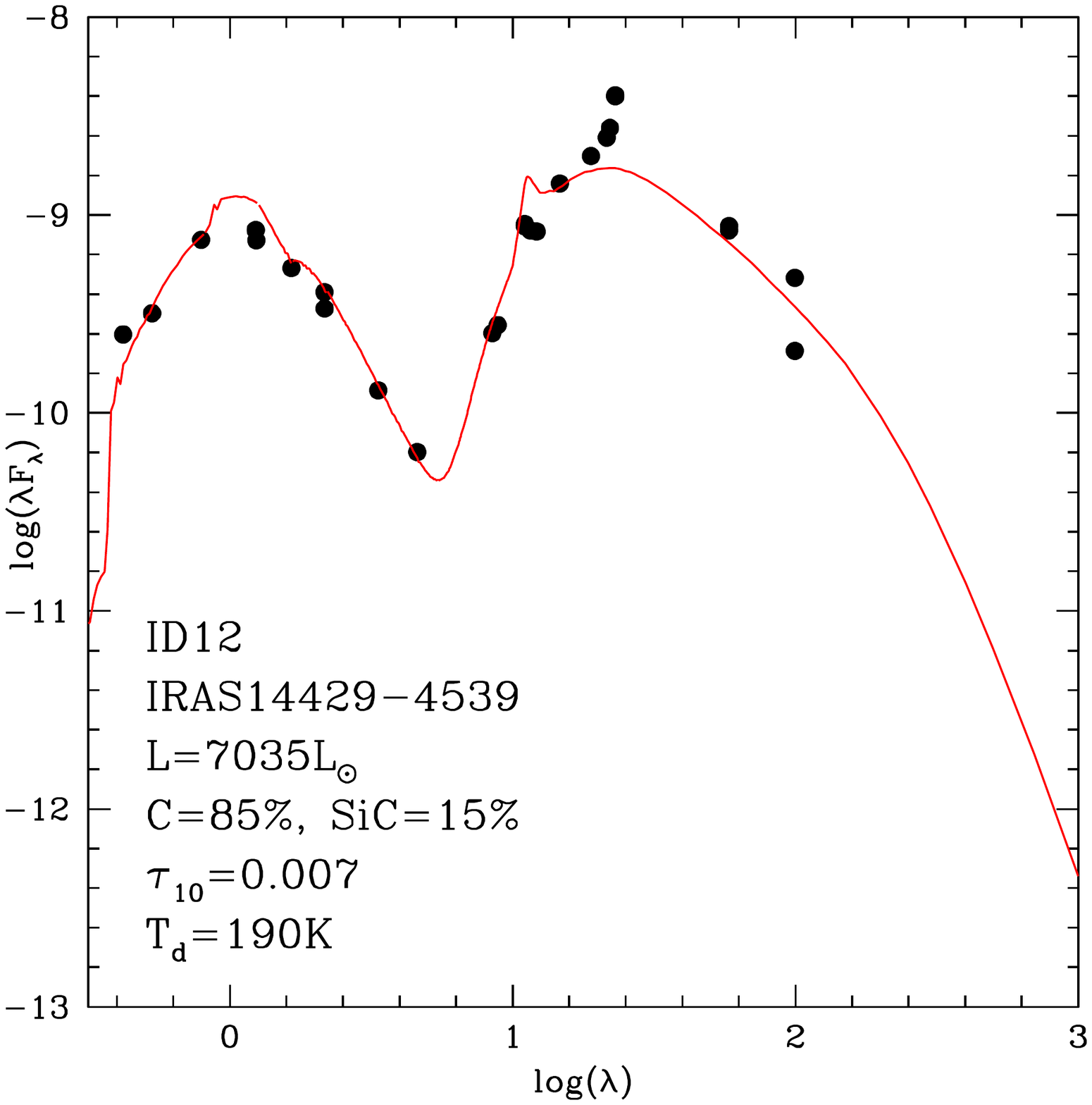}}
\end{minipage}
\vskip-60pt
\begin{minipage}{0.32\textwidth}
\resizebox{1.\hsize}{!}{\includegraphics{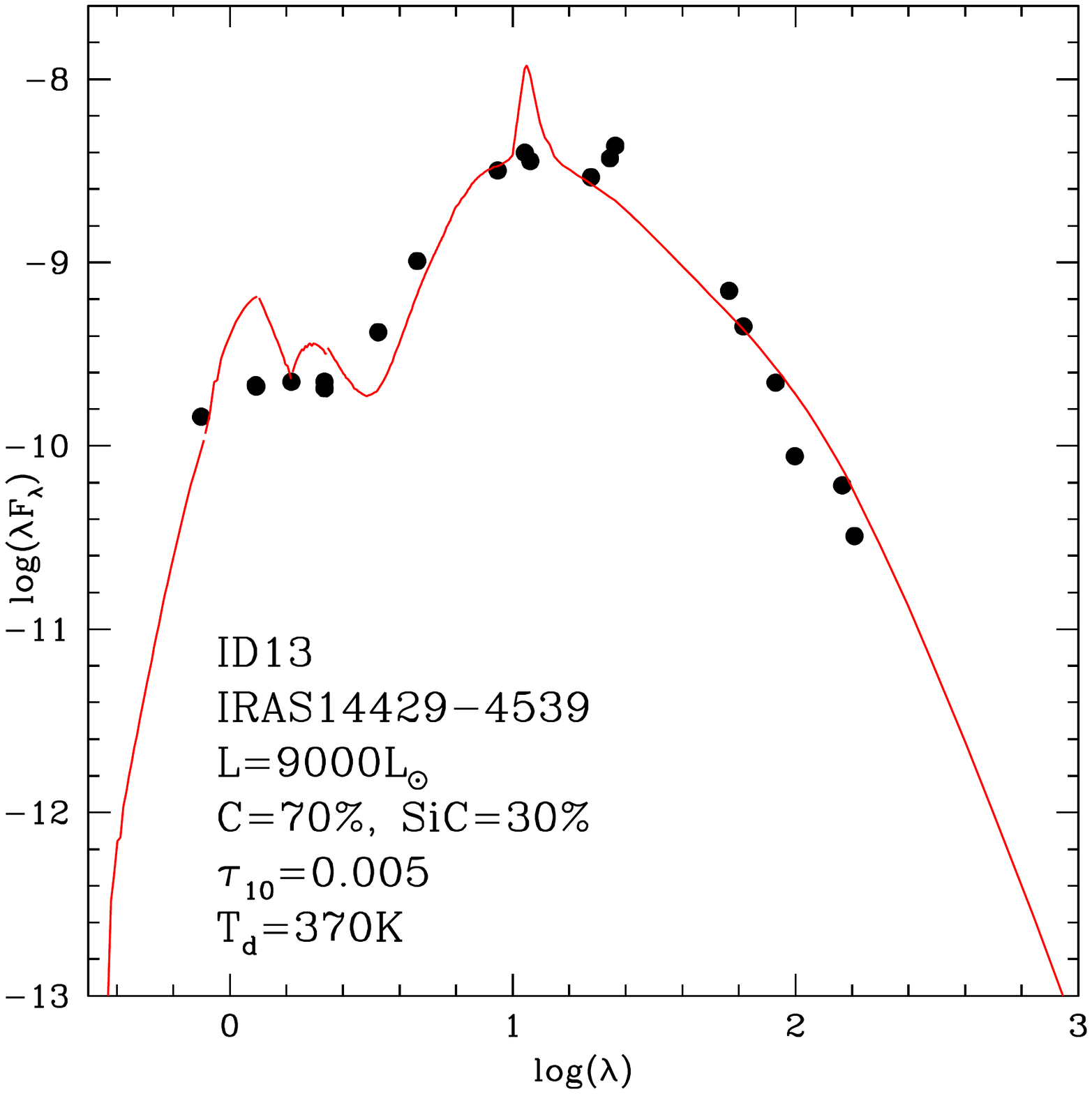}}
\end{minipage}\begin{minipage}{0.32\textwidth}
\resizebox{1.\hsize}{!}{\includegraphics{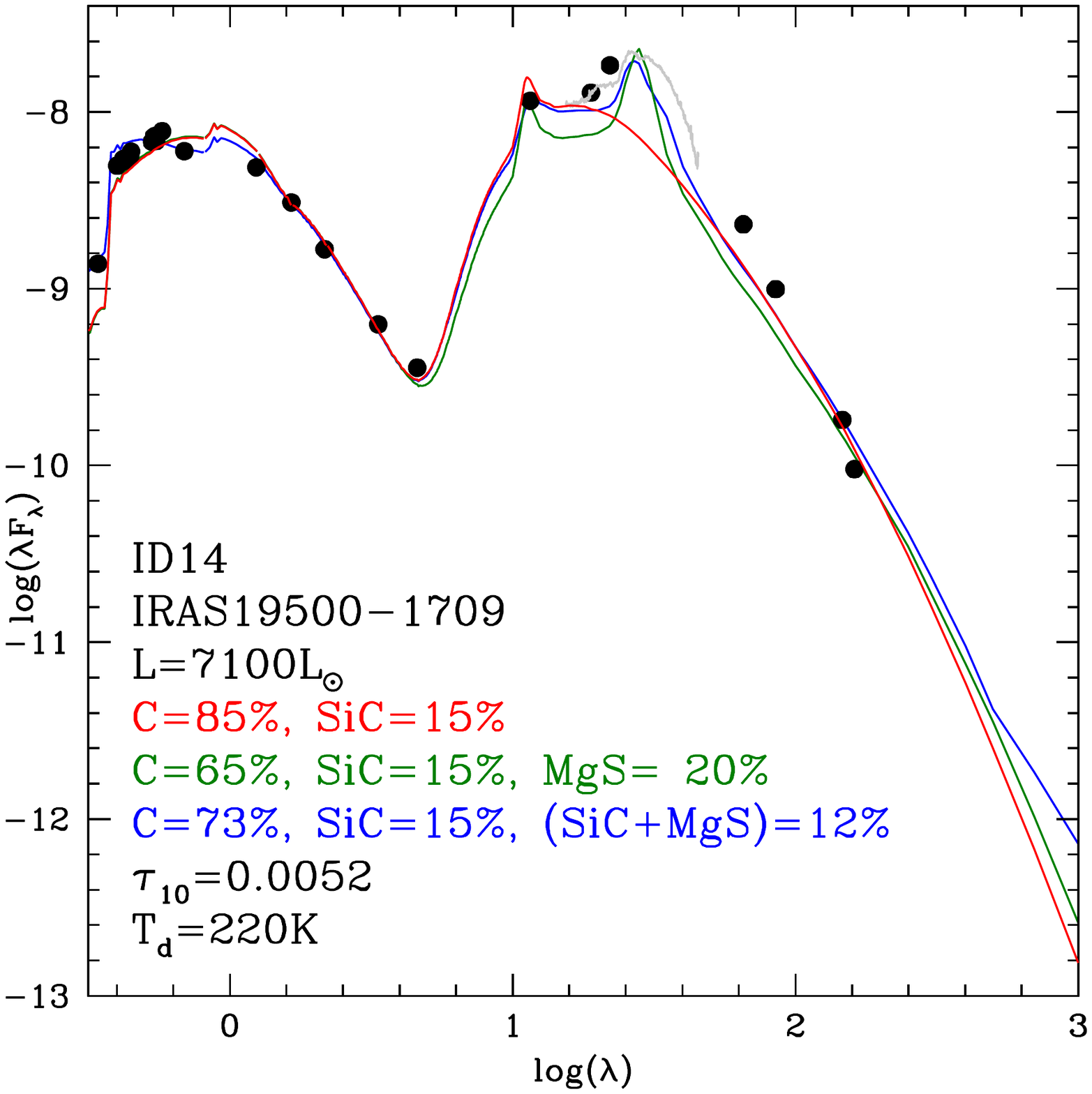}}
\end{minipage}
\begin{minipage}{0.32\textwidth}
\resizebox{1.\hsize}{!}{\includegraphics{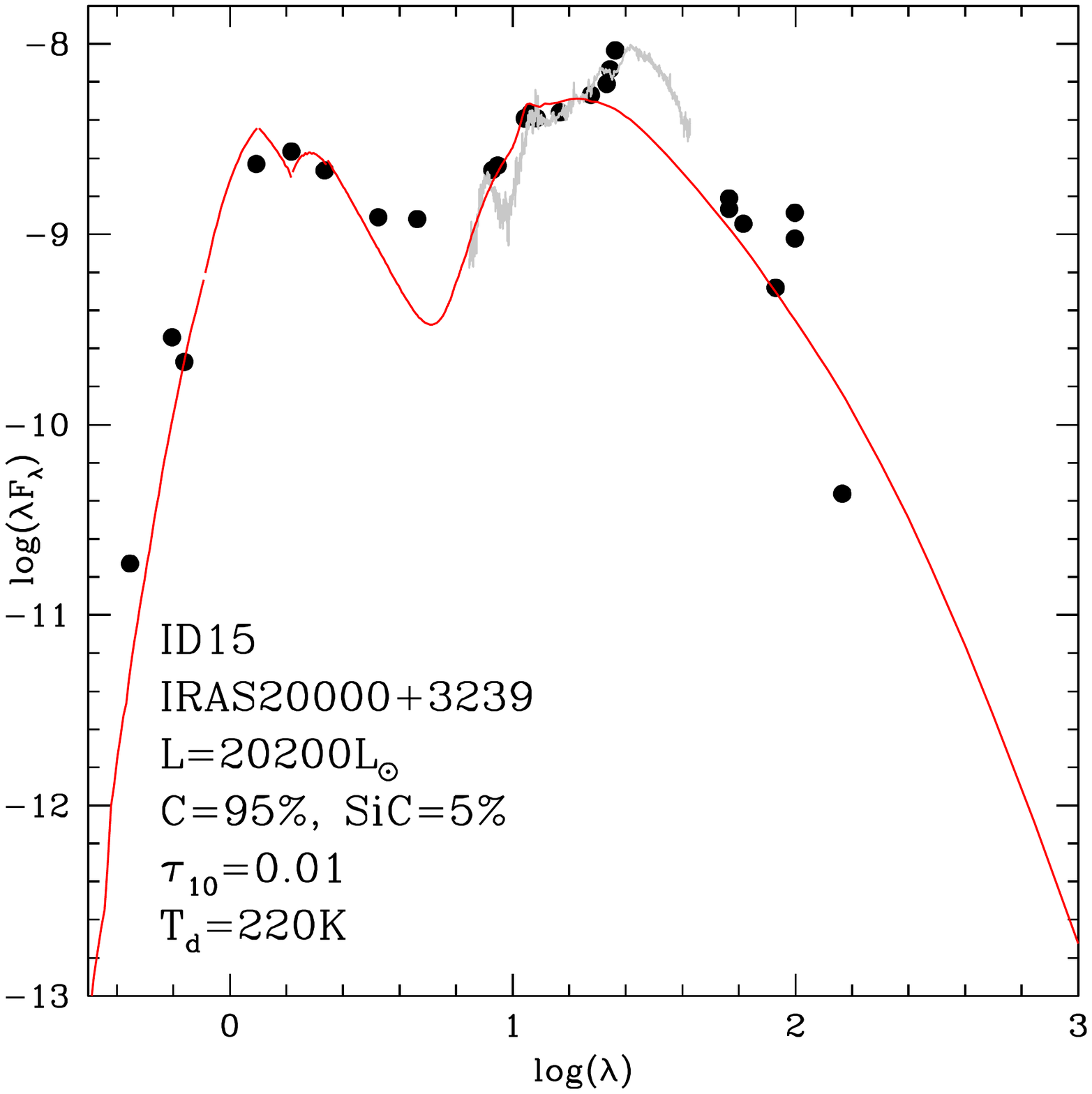}}
\end{minipage}
\vskip-60pt
\begin{minipage}{0.32\textwidth}
\resizebox{1.\hsize}{!}{\includegraphics{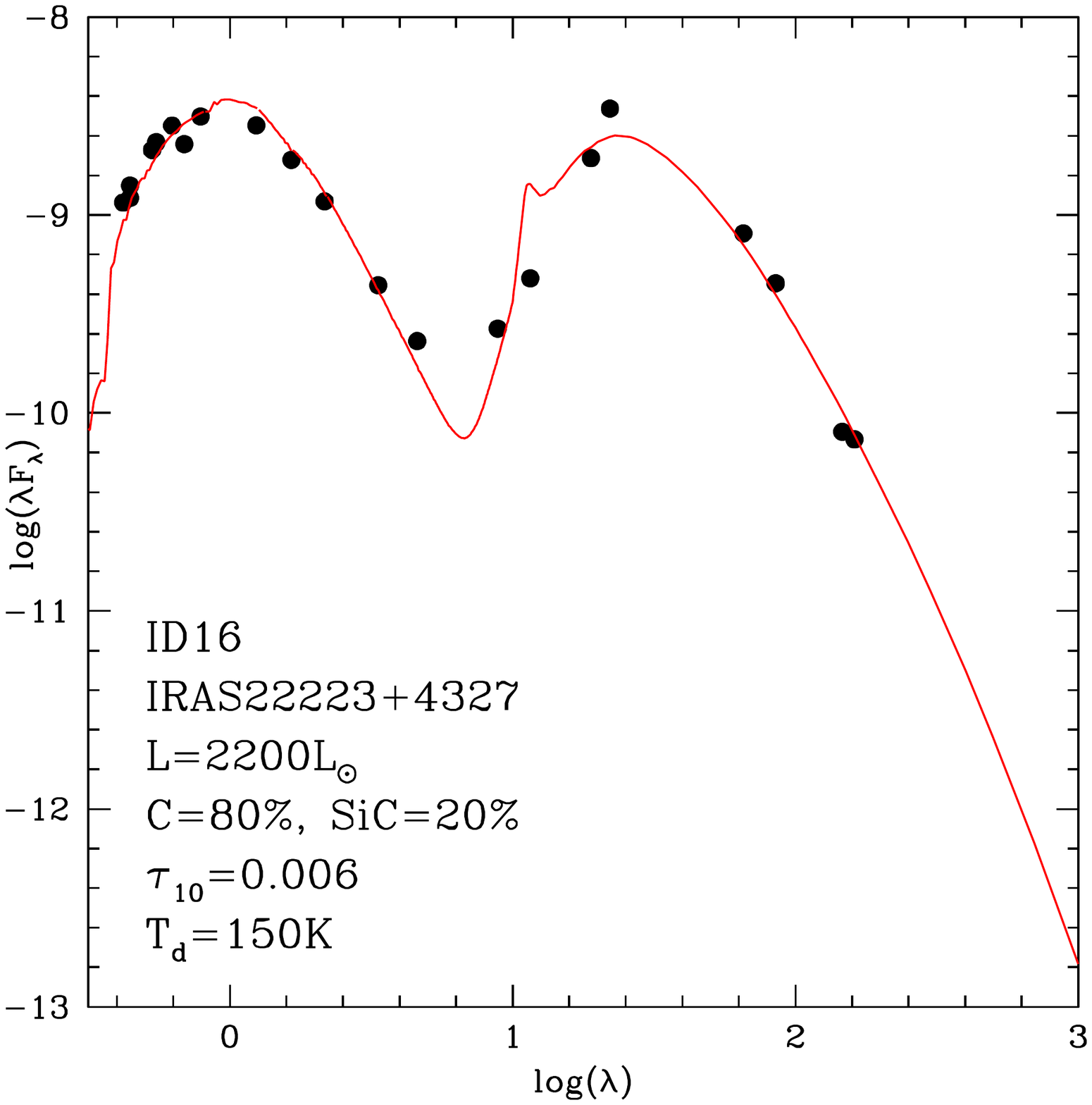}}
\end{minipage}
\begin{minipage}{0.32\textwidth}
\resizebox{1.\hsize}{!}{\includegraphics{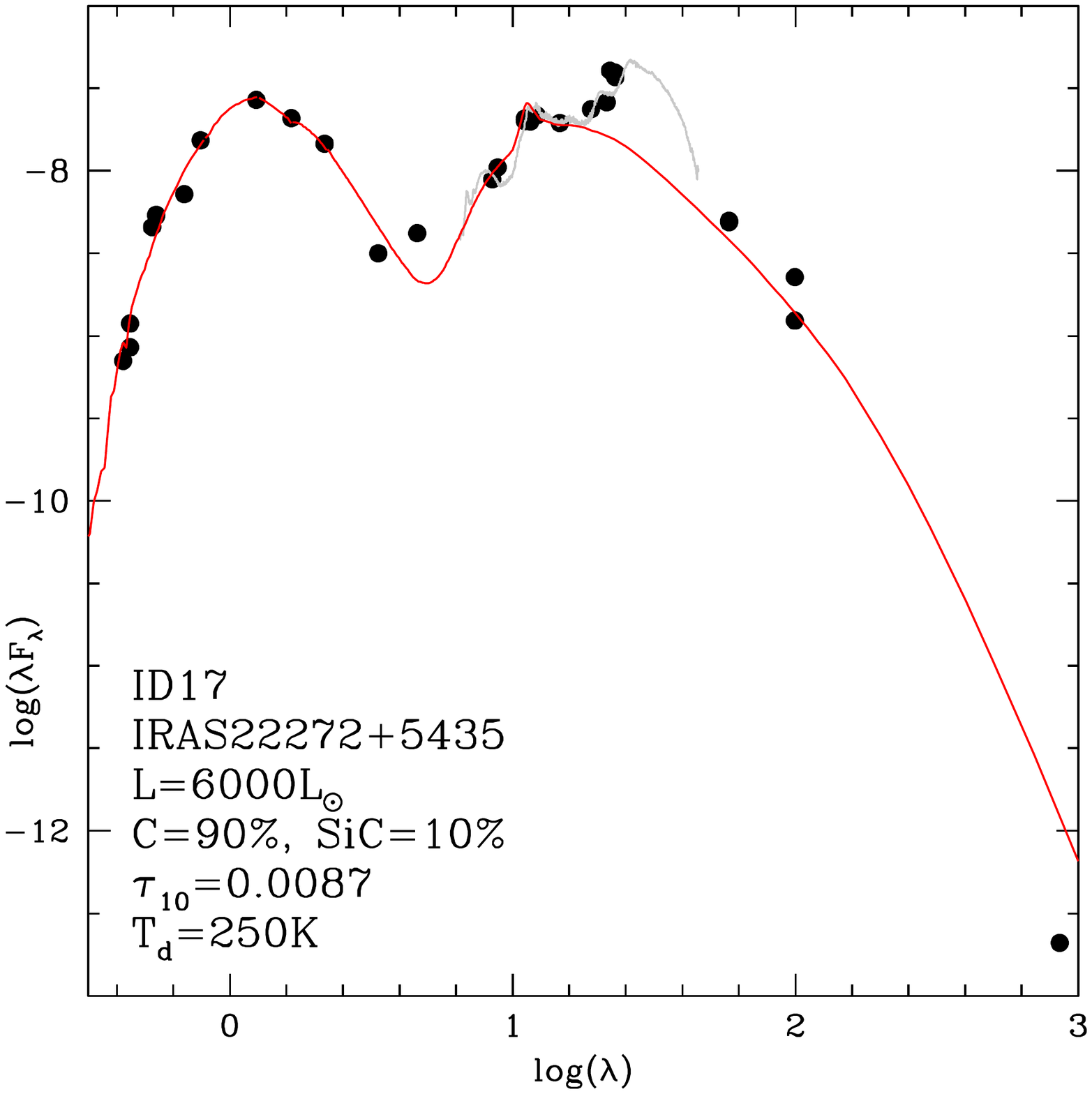}}
\end{minipage}
\begin{minipage}{0.32\textwidth}
\resizebox{1.\hsize}{!}{\includegraphics{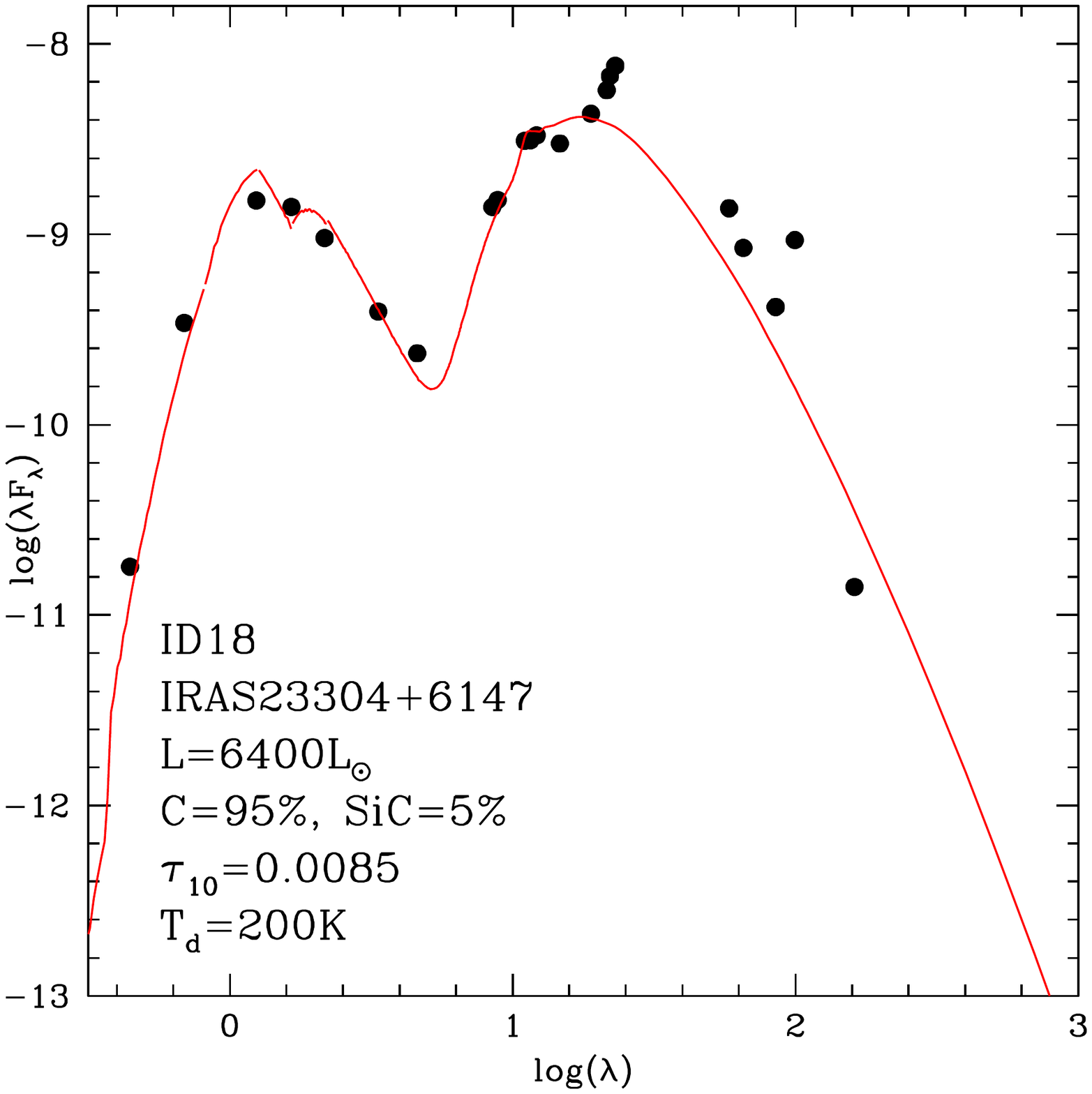}}
\end{minipage}
\vskip-40pt
\caption{Same as in Fig. \ref{fsed}. The different lines in the central
panel indicate the synthetic SED obtained by assuming a SiC plus solid carbon
dust mixtre (red line), $12\%$ of pure MgS dust (green line), $12\%$ of
hybrid dust with a SiC core and MgS mantle (blue line). The grey line shows the SWS spectra from \citet{sloan03}.}
\label{fsed2}
\end{figure*}

\section{The methodology to characterise the individual sources}
\label{method}The sample of 19 C-rich post-AGB stars showing $s$-process enhancement presented in K22 and
studied in K23 was investigated. A complete set of surface abundances of these stars was compiled by the authors of  K22. Further details on the individual objects and the surface abundance analysis are available \citep[e.g.][]{reddy99,reyneirs00,hans00,klochkova06,pereira11,desmedt16}. The authors of K22 classified all 19 objects as single stars based on results from long-term radial velocity monitoring programmes \citep[e.g.][see K22 for more details.]{hrivnak17}

Our aim was to characterise the C-rich sources and  infer 
the dynamical properties of the outflow during the period from the last release of dust to the present. For this purpose, we followed the same method used in T22 to study the MC post-AGB stars presented by \citet{devika14, devika15} and \citet{flavia22} to study the M-star Galactic sample presented in K22.

The radiative transfer code DUSTY \citep{nenkova99} was used to build synthetic SEDs that were compared with the observed shape of the SED of the individual stars, reconstructed on the basis of photometric results from K22 and, when available, the spectrum from the Short Wavelength Spectrometer (SWS) of the \textit{Infrared Space Observatory} \citep[ISO;][]{sloan03}. The synthetic SEDs start with the effective temperatures and metallicities given in K22, which are required to select the appropriate atmosphere models within the ATLAS9 library \citep{castelli03} from which the input radiation from the star entering the dusty regions is derived.

The DUSTY code models the reprocessing of the radiation emitted from single stars by dusty regions with spherical spatial distribution. This methodology can be confidently applied in the present context since the target sample consists of single sources (as classified in K22) with a shell-type SED indicative of a circumstellar environment with a shell-like dust distribution. In this work, DUSTY was used in the standard mode
where the density stratification is not derived by hydrodynamic calculations but is rather imposed a priori.
This approach is consistent with previous investigations on the expected SED from evolved stars where
the mass-loss rates and, consequently, the density stratification of the wind are given 
as input to be consistent with results from stellar evolution (which give the mass-loss rate)
and dust formation modelling (which provides the density profile of the wind).

As discussed in T22, the detailed comparison between the synthetic and the
observed SED of post-AGB stars allows us to derive the following: a) the mineralogy of the circumstellar dust, which in this specific case refers to the relative fractions of amorphous carbon (we used the ACAR optical constants by Zubko et al. 1996) and silicon carbide (SiC; optical constants by Pegourie 1988); b) the optical depth at the wavelength $\lambda=10~\mu$m, that is, $\tau_{10}$; and c)  the dust temperature at the inner edge of the dusty shell, ${\rm T}_{\rm d}$, which is tightly correlated to the distance of the inner border of the dusty zone from the centre of the star, ${\rm R}_{\rm in}$. The typical uncertainties associated with these estimates, as described in Section 3 of T22, are approximately 10\% in the fraction of SiC, 10-15\% in $\tau_{10}$,  and $20$K for ${\rm T}_{\rm d}$.

The authors of T22 discussed the possibility of deriving the luminosity of the stars by reproducing the near-IR spectrum. Determining the luminosity requires a distance, which is known for stars in the MC. For Galactic sources, distances can be estimated from $\it{Gaia}$ DR3 parallaxes.  We used the probabilistic distances from \citet{bailer-jones2021}. 
The authors of K22 reported the range of luminosities based on these distances (see their Table 1, which also also gives the renormalised unit weight error (RUWE)\footnote{The RUWE is a goodness-of-fit statistic from the astrometric procedure used to determine positions, parallaxes, and proper motions of sources observed by Gaia \citep{lindegren2021}. High RUWE values are an indication of a poor astrometric fit (which might be due to unresolved multiplicitiy, \citep[e.g. ][]{belokurov2020, penoyre2020}); and a RUWE value of less than1.4 is typically employed to select stars with an accurate astrometric solution.} of each star). 
We adopted the same classification used in K22 where the stars were flagged as either Q1 or Q2 if the RUWE was below or above 1.4, respectively. 

For Q1 stars, the luminosity ranges were found by fitting the SEDs (in particular by scaling the synthetic SED to match the near-IR spectrum) combined with the distance range given in K22. From these ranges, we selected the values most consistent with the interpretation of the individual sources from K23. 
For the Q2 sources, we first derived the luminosity by combining results from the SED fitting with the average distances given in K22. Unlike the Q1 case, for Q2 stars we also considered distances beyond the range used in K22 in order to make the observations and the modelling consistent. This procedure was adopted in particular for three objects, discussed in Section \ref{hbstars}, for which we invoke distances approximately two to 20 times higher than those given in K22, which
is justified considering the large RUWE index of these objects.

The following sections focus on the IR excess of the stars, which we used to determine the evolutionary phase when the dust formed, the evolution of the physical conditions of the star, and the efficiency of the dust formation process from the tip of the AGB (TAGB). We use `TAGB' to indicate the evolutionary phase where the radius of the star is the greatest before the beginning of the contraction towards the post-AGB evolution until the present epoch. 

We relied on the results from stellar evolution calculations to model in detail the late-AGB and post-AGB evolutionary phase. The authors of K23 presented these results obtained with the ATON code \citep{ventura98} for almost all the stars in the sample, with a few exceptions that will be addressed in detail in Section \ref{id817}.

We considered the hypothesis that the dust was released in various evolutionary phases during the post-AGB evolution. For each of the selected stages, we modelled dust formation following the scheme introduced by the Heidelberg group \citep{fg06}, which has already been applied in previous works by our team \citep[][T22]{ventura12, ventura14}. We checked the consistency between the results obtained from dust formation modelling and SED fitting by considering the two following points.

First, for each of the evolutionary phases considered, we modelled
dust formation to calculate the corresponding optical depth 
$\tau_{10}^{\rm onset}$, which
would characterise the star if it was observed in that epoch. To check 
compatibility with the current optical depth derived from SED fitting,
$\tau_{10}^{\rm now}$, we applied the following scaling relation
introduced in T22:
$$\tau_{10}^{\rm now}=\tau_{10}^{\rm onset}\times {{\rm R}_{\rm dust}\over{{\rm R}_{\rm in}}}
,\eqno{(1)}$$
where ${\rm R}_{\rm dust}$ is the distance of the dusty region from the 
centre of the star at the time when the dust was produced (typically five to ten
stellar radii from the centre of the star), while ${\rm R}_{\rm in}$
is the current distance of the dust from the post-AGB star.
We evaluated the reliability of the hypothesis that the dust 
was released at a given evolutionary phase by comparing $\tau_{10}^{\rm now}$
as derived from Eq.~1 with the values derived from the SED fitting, reported
in Table \ref{tabero}.

Then, to find the consistency, we followed the methodoloby of T22 by identifying the evolutionary
stage (hence, the effective temperature considering that the luminosity
is approximately constant) during which the dust responsible for the  detected IR excess was released. Therefore, we derived the velocity
with which the wind would move since the dust was released until 
the present epoch by dividing the derived ${\rm R}_{\rm in}$ by the time interval
required for the effective temperature of the star to increase from the value
at the considered evolutionary phase until the current value. The latter
time was calculated based on stellar evolution modelling. We considered
only the evolutionary phases for which the derived velocities are in
the $5-20$ km$/$s range, which reflected an uncertainty onto the 
effective temperature of approximately 500 K.

By combining the analyses from the two described points, we could estimate the mass-loss rate at the TAGB, 
the rate at which $\dot{\rm M}$ decreases while the stars contract to the post-AGB phase, and
the phase when the dust observed at the present epoch was formed. We discuss an application of this 
methodology to the study of some sources in Section \ref{id817}.

\section{SED fitting}
\label{sedfit}

The results of the analysis performed on each of the objects in the sample are presented in Figs.~\ref{fsed} and \ref{fsed2}. 
In the cases where multi-epoch observations for the same band are available, all the results are shown in a vertical sequence according to the central wavelength of the corresponding filter. The figures show the SWS spectra from \citet{sloan03}, 
when available. 

The synthetic spectra shown in Fig.~\ref{fsed} and \ref{fsed2} do not reproduce the SWS spectra in the spectral regions 
around the features at $6.9~\mu$m, $15-21 \mu$m, and $30~\mu$m. Regarding the latter feature, which is by far the strongest, Goebel and Moseley (1985)
proposed that it is originated by the presence of MgS particles in the surroundings of C-stars.
We tested the possibility that either pure MgS or MgS dust coated on SiC particles are
responsible for the feature. The results show that the estimated
luminosity and overall optical depth of the individual sources are scarcely affected  by the inclusion of 
MgS particles because 
the luminosity, as discussed in T22, is derived from the mid-IR flux, whereas the optical
depth is almost entirely due to the solid carbon dust. The features at $6.9~\mu$m and at $15-21 \mu$m
were not included in the simulations, as the carriers are unknown. However, since
these features are weaker than the $30~\mu$m feature, their inclusion in synthetic modelling
would not change the conclusions that have been reached.

Accounting for the presence of MgS dust, as discussed in the Appendix, leads to a better agreement between the synthetic and the SWS spectra. This result further strengthens the
possibility that MgS particles (likely coated on SiC grains, see the Appendix) are the carriers
responsible for the  $30~\mu$m feature. However, further investigations based on updated
optical constants for MgS dust and on a wider sample of post-AGB sources are
required before definitive conclusions can be reached.

The quantities reported in the different panels of Fig.~\ref{fsed2} were found following the method described in 
the previous section. We observed a wide spread in luminosity, which extends up to $\sim 20000~{\rm L}_{\odot}$,
and in optical depths, as the stars with the largest IR excess were characterised
by $\tau_{10} \sim 0.025$. In the SED of some stars we recognised the presence of SiC particles,
in percentages between 0 and $30\%$, as the only way of reproducing
the sharp feature at $11.3~\mu$m. The fraction of SiC can only be reliably assessed amongst stars for which the spectrum is available. Unfortunately in some cases, this feature is partly 
overlapped with the one associated with the aforementioned complex hydrocarbons, complicating analysis. 

Table~\ref{tabero} reports a summary of the results obtained from the SED fitting. The table
lists the metallicities and the effective temperatures derived spectroscopically in K22, 
the luminosities derived in the present study, and the dust properties (i.e. the mineralogy, the 
optical depth, the dust temperature, and the distance from the centre to the star of the inner 
border of the dust layer). The last column reports the quality flag of the distance as given in
K22.

\begin{table*}
\caption{Physical and dust properties of the Galactic post-AGB stars in this study.
The columns are defined as follows (from left to right):
Column 1: Object name; 2: Source ID (from K22); 3-4: Metallicity and effective temperatures as listed in K22 (see their Table 1); 5: Luminosity derived in the present study (see Section~\ref{method}); 6: Luminosity from K22 (see their Table 1), obtained considering the upper and lower limits of the distances retrieved from \citet{bailer-jones2021}; 7: C/O from K22 (see their Table 3); 8-10: Optical depth at $10~\mu$m, dust temperature, and distance separating the central star from the inner border of the dusty region, found via SED
fitting; 11: The quality flag (Q1 or Q2), based on the Gaia EDR3 RUWE parameter
from K22. The last eight rows are the carbon stars in the sample of T22. 
}
\label{tabero}      
\centering
\addtolength{\leftskip}{-0.8cm}
\begin{tabular}{l c c c c c c c c c c c c c c}    
\hline      
Source & ID & $[$Fe$/$H$]$ & ${\rm T}_{\rm eff}$[K] & ${\rm L}/{\rm L}_{\odot}$ & ${\rm L}/{\rm L}_{\odot}^{K22}$ & $\rm{C}/\rm{O}$ & $\tau_{10}$ & ${\rm T}_d$[K] & ${\rm R}_{\rm in}/{\rm R}_{\odot}$ & flag  \\
\hline 
\\
IRAS Z02229+6208 & 1 & $-0.45\pm 0.14 $ & 5952  & 13000 & 9973 -- 1611  & $-$             & 0.0082 & 300 & $1.84 \times 10^5$ & Q2 \\
IRAS 04296+3429 & 2  & $-0.62\pm 0.11$  & 7272  & 6000  & 5971 -- 20082 & $-$             & 0.007 & 280 & $1.74 \times 10^5$ & Q2  \\
IRAS 05113+1347 & 3  & $-0.49\pm 0.15$  & 5025  & 2100  & 1043 -- 6731  & $2.42\pm 0.40$  & 0.026  & 200 & $1.84 \times 10^5$ & Q2 \\
IRAS 05341+0852 & 4  & $-0.54\pm 0.11$  & 6274  & 300   & 197 -- 592    & $1.06\pm 0.30$  & 0.028  & 320 & $2.60 \times 10^4$ & Q2 \\
IRAS 06530-0213 & 5  & $-0.32\pm 0.11$  & 7809  & 6900  & 2736 -- 8178  & $1.66\pm 0.36$  & 0.0047 & 200 & $4.64 \times 10^5$ & Q2 \\
IRAS 07134+1005 & 6  & $-0.91\pm 0.20$  & 7485  & 5500  & 4955 -- 6098  & $1.24\pm 0.29$  & 0.006  & 210 & $3.68 \times 10^5$ & Q1 \\
IRAS 07430+1115 & 7  & $-0.31\pm 0.15$  & 5519  & 20    & 14 -- 30      & $1.71\pm 0.30$  & 0.020  & 230 & $1.39 \times 10^4$ & Q2 \\
IRAS 08143-4406 & 8  & $-0.43\pm 0.11$  & 7013  & 5400  & 3927 -- 5452  & $1.66\pm 0.39$  & 0.002  & 150 & $6.81 \times 10^5$ & Q1 \\
IRAS 08281-4850 & 9  & $-0.26\pm 0.11$  & 7462  & 7900  & 5567 -- 16692 & $2.34\pm 0.24$  & 0.007  & 280 & $1.94 \times 10^5$ & Q2 \\     
IRAS 12360-5740 & 10 & $-0.40 \pm 0.15$ & 7273  & 6300  & 5178 -- 7940  & $0.45\pm 0.20$  & 0.0035 & 180 & $4.81 \times 10^5$ & Q1 \\
IRAS 13245-5036 & 11 & $-0.30 \pm 0.10$ & 9037  & 7100  & 7106 -- 16800 & $1.11\pm 0.30$  & 0.004  & 220 & $3.89 \times 10^5$ & Q2 \\
IRAS 14325-6428 & 12 & $-0.56\pm 0.10$  & 7256  & 7035  & 3758 -- 6988  & $2.27\pm 0.40$  & 0.007  & 190 & $5.22 \times 10^5$ & Q2 \\
IRAS 14429-4539 & 13 & $-0.18\pm 0.11$  & 9579  & 9000  & 1591 -- 14624 & $1.29\pm 0.26$  & 0.0085 & 360 & $1.20 \times 10^5$ & Q2 \\
IRAS 19500+1709 & 14 & $-0.59\pm 0.10$  & 8239  & 7100  & 6194 -- 8138  & $1.02\pm 0.17$  & 0.0052 & 220 & $3.73 \times 10^5$ & Q1 \\
IRAS 20000+3239 & 15 & $-1.40 \pm 0.20$ & 5478  & 20200 & 9332 -- 25218 & $-$             & 0.010  & 220 & $5.32 \times 10^5$ & Q2 \\
IRAS 22223+4327 & 16 & $-0.30 \pm 0.11$ & 6008  & 2200  & 1956 -- 2499  & $1.04\pm 0.22$  & 0.006  & 150 & $4.93 \times 10^5$ & Q2 \\
IRAS 22272+5435 & 17 & $-0.77\pm 0.12$  & 5325  & 6000  & 5234 -- 6108  & $1.46\pm 0.26$  & 0.0087 & 250 & $1.90 \times 10^5$ & Q1 \\
IRAS 23304+6147 & 18 & $-0.81\pm 0.20$  & 6276  & 6400  & 6381 -- 9386  & $2.8\pm 0.2$    & 0.0085 & 200 & $4.22 \times 10^5$ & Q2 \\
\\
\hline
\\
J052220.98  & LMC & -0.50  & 5750 & 4500  & -- &  0.015 & 220 & -- & $2.63\times 10^5$  \\
J053250.69  & LMC & -1.10  & 6000 & 5200  & -- &  0.006 & 250 & -- & $2.06\times 10^5$  \\
J004114.10  & SMC & -1.04  & 5750 & 5800  & -- &  0.007 & 280 & -- & $1.65\times 10^5$  \\
J050632.10  & LMC & -0.40  & 7600 & 6000  & -- &  0.003 & 240 & -- & $2.74\times 10^5$  \\
J003643.94  & SMC & -0.63  & 7500 & 6500  & -- &  0.002 & 250 & -- & $2.56\times 10^5$  \\
J051848.84  & LMC & -1.00  & 6000 & 6700  & -- &  0.008 & 240 & -- & $2.62\times 10^5$  \\
J004441.03  & SMC & -1.07  & 6000 & 8500  & -- &  0.002 & 320 & -- & $6.64\times 10^4$  \\
J005803.08  & SMC & -1.03  & 6500 & 12000 & -- &  0.025 & 280 & -- & $1.20\times 10^5$  \\
\hline
\end{tabular}
\end{table*}

\section{Physical, dust, and optical properties of the stars}
\label{disc}
To characterise the individual sources in terms of the dust currently around them and to find the appropriate correlation between the current IR
excess and the previous history of the stars, the authors of T22 proposed studying the correlation
between the optical depth $\tau_{10}$ derived from SED fitting with the luminosity (L) of the
individual stars and the distance ($\rm R_{\rm in}$) of the inner border of the 
dusty region from the stellar surface. Fig.~\ref{fsumm} presents the $\tau_{10}-{\rm L}$ and $\tau_{10}-{\rm R}_{\rm in}$  
trends defined by the stars investigated in the present sample. In the following paragraphs, we discuss the different groups of stars divided according to the
mass and formation epoch of their progenitors and by the amount of dust present
in their surroundings.

\subsection{Low-mass carbon stars}
\label{lowmc}
The most populated region in the $\tau_{10}-{\rm L}$ plane has
luminosities in the $5000-10000~{\rm L}_{\odot}$ range and optical depths of
$\tau_{10}<0.01$. The authors of K23 interpreted the sources in this region as being the descendants
of $0.9~{\rm M}_{\odot} < {\rm M} < 1.5~{\rm M}_{\odot}$ stars that reached the C-star stage towards their final evolutionary stages, after the initial AGB lifetime, during which they evolved as M-type stars.

The distribution of the stars in this plane shows that when stars of similar
metallicity are considered, the optical depth generally increases with 
luminosity. This is consistent with the analysis presented in T22, which showed that the brighter post-AGB carbon stars in the MCs are characterised
by larger IR excesses. The authors in that paper suggested that this is partly 
due to the fact that the brighter carbon stars experience higher dust formation 
rates during the final AGB phases and that higher-luminosity
stars evolve faster towards the post-AGB stage. Thus, the time elapsed since
the dust was released until the present epoch is shorter, and the dusty region is
currently closer to the surface of the star. This interpretation is confirmed by the distribution of the same stars in the 
$\tau_{10}-{\rm R}_{\rm in}$ plane, where with the exceptions of
the sources ID 15 and ID 18 (both flagged as Q2), the distance of the 
dusty region from the surface of the star is negatively correlated
with the luminosity.

\begin{figure*}
\begin{minipage}{0.48\textwidth}
\resizebox{1.\hsize}{!}{\includegraphics{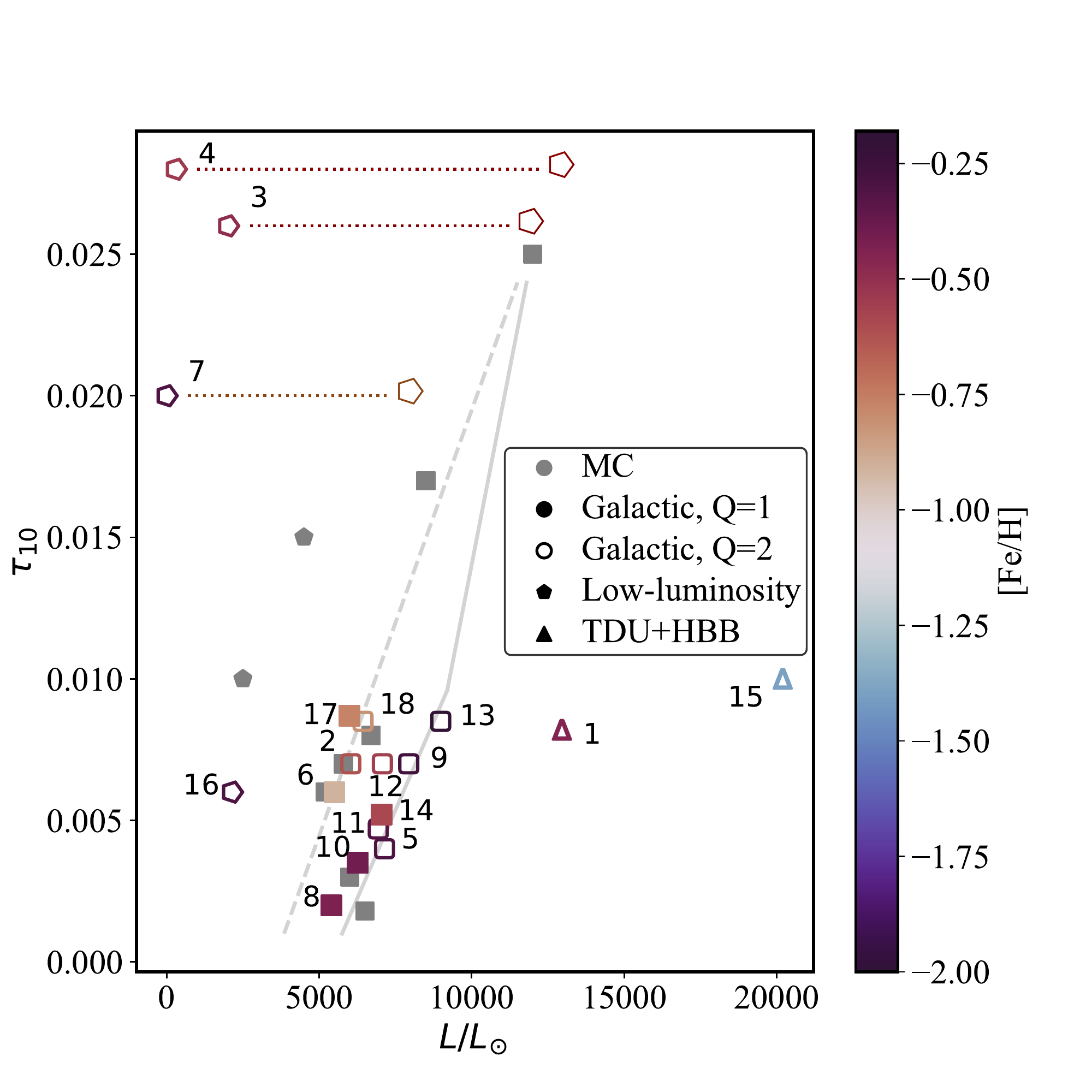}}
\end{minipage}
\begin{minipage}{0.48\textwidth}
\resizebox{1.\hsize}{!}{\includegraphics{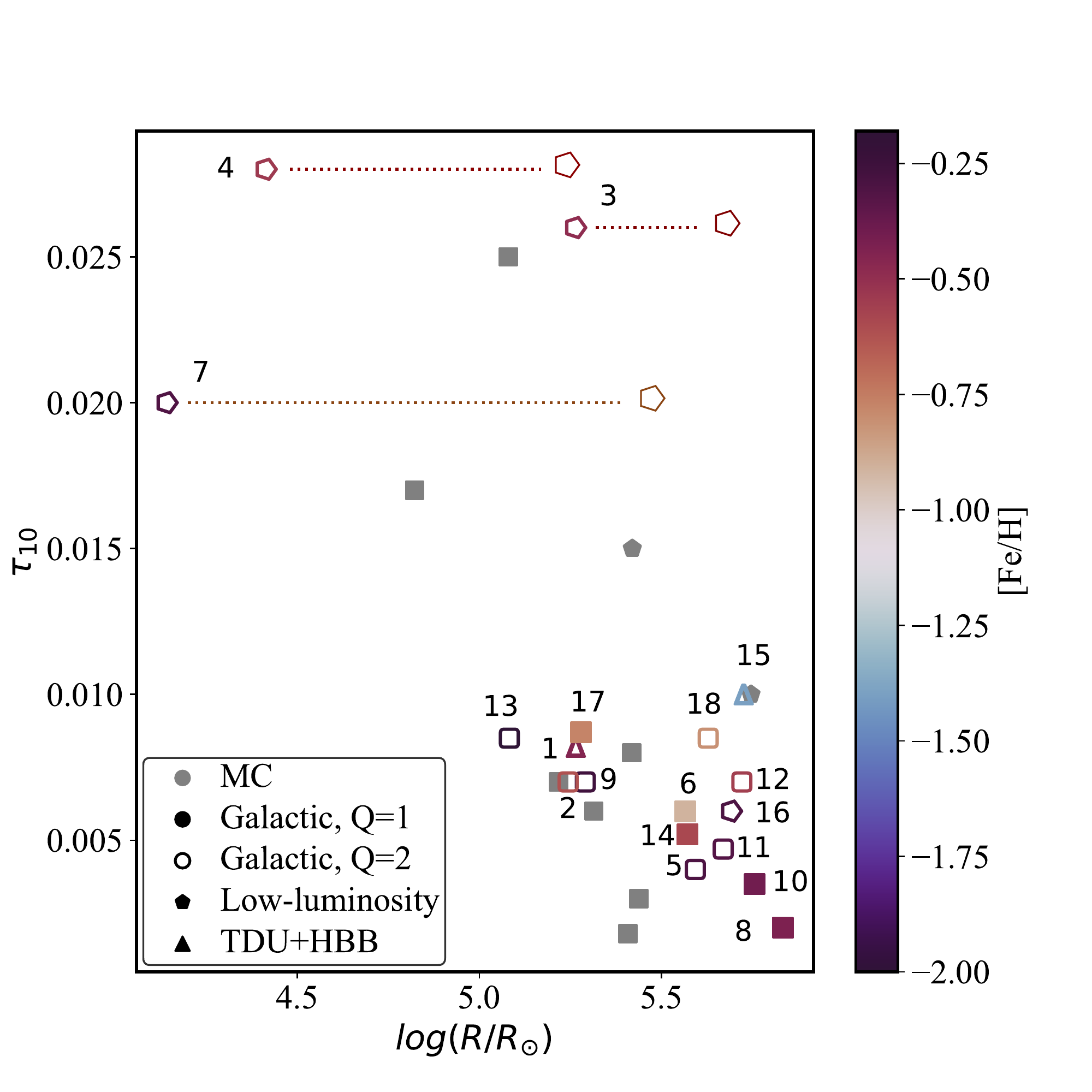}}
\end{minipage}
\caption{Optical depths at $10~\mu$m as a function of the luminosity of the star (left panel) and of the distance of the inner border of the dusty zone from the centre of the star (right panel). The physical and dust parameters are derived from SED fitting (see Section~\ref{sedfit}). The vertical colour bar indicates the 
relative metallicity of the sources. The Galactic post-AGB stars flagged as Q1 are reported with filled symbols, while those flagged as Q2 are reported with empty symbols. The grey markers refer to the LMC stars studied by T22. The dotted lines show the projection of the stars obtained by changing the distance retrieved from \citet{bailer-jones2021} and therefore the brightness. Table \ref{tabero} reports the uncertainties in luminosity. In the left panel, the dashed grey line represents the trend in the $\tau-$L plane for the metal-poor stars of the Galaxy and the MC, and the solid grey line represents the trend in the $\tau-$L plane for the metal-rich objects of the same sample.   }
\label{fsumm}
\end{figure*}

A further result visible in the left panel of Fig.~\ref{fsumm} is that the
stars of different metallicities define separate sequences, with
metal-poor objects characterised by a more prominent IR
excess (hence higher optical depths) compared to their more metal-rich
counterparts of similar luminosity. The more metal-rich sequence extends
from luminosities of $\sim 5000~{\rm L}_{\odot}$ and $\tau_{10} \sim 2\times 10^{-3}$
(this is represented by the source ID 8) to ${\rm L}\sim 9000~{\rm L}_{\odot}$
and $\tau_{10} \sim 8\times 10^{-3}$ (ID 13). The metal-poor star sequence
can be ideally linked to the two bright metal-poor carbon stars studied in
T22 and easily identified in Fig.~\ref{fsumm}. The luminosities
in this case range from $5000$ to $12000~{\rm L}_{\odot}$, and the variation of the optical depth
is $6\times 10^{-3} < \tau_{10} < 2.5\times 10^{-2}$.

The higher opacity of metal-poor stars has two causes that are closely
correlated with each other. As a first approximation, if we consider low-mass
stars of similar luminosity and different chemical composition, the surface
abundance of $^{12}$C upon entering the post-AGB phase does not depend on metallicity. On the other hand, metal-poor stars achieve
a higher surface C/O ratio than their more metal-rich counterparts of
similar luminosity because the initial oxygen in the star is lower.
This results in higher surface molecular opacities and thus more expanded and
lower surface gravity structures, which favour a cooler and denser
envelope. Hence, the environment is more favourable to the formation of dust.

An additional cause of the higher IR excess of metal-poor stars
is that the key quantity driving the formation of carbon dust is the carbon
excess with respect to oxygen \citep{fg02, fg06}, which is intrinsically higher in metal-poor
stars, again due to the lower oxygen content. This characteristic causes a higher quantity of
carbon molecules to be available for the formation of solid carbon grains. \citet{lagadec08} and \citet{sloan12} reached similar conclusions.

Luminosity also plays a role in this context, as brighter stars 
experience higher carbon enrichment at the surface before the envelope
is lost. This is not only because their envelope is more massive than their lower
luminosity counterparts, but also because the inter-pulse periods decrease
with luminosity. Thus, for a given mass-loss rate, a smaller amount of mass is lost between the two thermal pulses (TP) that follow. 
At $L \sim 10000~{\rm L}_{\odot}$, the amount of carbon
accumulated in the surface regions (above $1\%$ mass fraction) becomes much higher than oxygen and
renders the (C-O) excess substantially independent of metallicity. Therefore, the 
separation in terms of $\tau_{10}$ between metal-poor stars and their higher metallicity counterparts is 
reduced as the
luminosity increases. Therefore, post-AGB carbon stars
with luminosities of around $10000~{\rm L}_{\odot}$ should populate the same
regions of the plane hosting the two most opaque stars in the LMC studied by
T22 independent of the metallicity.

\subsection{Stars experiencing hot bottom burning and third dredge-up}
The stars ID 1 and ID 15 populate the upper-right region of the
$\tau_{10}-{\rm L}$ plane, with luminosities of$15000-20000~{\rm L}_{\odot}$
and optical depths of $\tau_{10} \sim 0.01$. The authors of K23 identified these stars
as being the progeny of $3-3.5~{\rm M}_{\odot}$ stars formed around 300 Myr ago
whose surface chemical composition was altered by the combined action of
HBB and TDU. Considering ID 15 and its low metallicity, however, this understanding is partly in tension with Galaxy evolution theories, 
according to which state that poor formation of metal-poor stars took place in recent 
epochs. Thus, ID 15 would be among the minority of objects formed during this
period. The interpretation for ID 1 and ID 15 given in K23 was based on the large
luminosity of these sources, which are close to the threshold required for the
ignition of HBB, and on the large nitrogen enhancement, a clear
signature of CN (or CNO) processing. Indeed, the stars whose surface
chemistry is affected by both HBB and TDU are those expected to experience
the largest N enrichment, as nitrogen is synthesised not only by the
$^{12}$C initially present in the star but also by the primary $^{12}$C
convected to the surface convective regions by the repeated TDU events.

 The sources ID 1 and ID 15 diverge from  the $\tau_{10}$ versus luminosity
trend defined by the lower mass counterparts discussed earlier in this section
since their optical depths are smaller than the two brightest
carbon stars in the LMC studied in T22. This is because the carbon excess with
which they enter the post-AGB phase is lower, owing to the effects
of HBB, which reduced the amount of $^{12}$C available in the surface
regions of the stars (see K23 for further discussion of the expected
evolution of the surface $^{12}$C in these stars).

The dust cloud around ID 1 is at $\sim 2\times 10^5~{\rm R}_{\odot}$. As Fig.\ref{fsumm} shows, this is one of the smallest dust distances in the sample. Because ID 1 is one of the brightest sources, this 
is once again consistent with the suggestion in T22 that luminosity 
plays a role in the evolutionary timescales. In fact, the transition of ID 1 from the late 
AGB to the post-AGB phase was significantly shorter than the transition for most of the 
carbon stars in our sample. 

The situation for ID 15 is more complicated, as we observed that the dusty
region is located at $\sim 6\times 10^5~{\rm R}_{\odot}$,
which is one of the largest distances in the sample (see Fig.\ref{fsumm})
even though ID 15 is the brightest star. This suggests that radiation
pressure might also have a role in this context.
For example, very bright stars experience higher radiation pressures than their
lower luminosity counterparts, and these higher pressures trigger faster winds.

\subsection{Post-HB or bright carbon stars?}
\label{hbstars}
The sample of stars considered in this work is completed by the sources 
ID 3, ID 4, ID 7, and ID 16, whose luminosities are significantly
below the threshold required for TP.
The authors of K23 suggested that these sources might be post-HB
objects that failed to reach the AGB phase. These objects may have lost their convective envelope and started the general contraction 
before reaching the AGB phase. Following the work of \citet{schwab20}, the argument presented
in K23 to explain the carbon and $s$-process enrichment exhibited
by these stars is that they experienced deep mixing during the
helium flash episode.

The present analysis poses a further problem for an exhaustive
understanding of the origin of these stars because, as clearly shown in
the left panel of Fig.\ref{fsumm}, all four of the least luminous stars are 
characterised by high optical depths and are among the highest in the 
sample. If these luminosities are confirmed, we should conclude 
that low-mass stars experience a short, intense episode of mass loss
before evolving to the blue side of the Hertzsprung–Russell (HR) diagram with the development 
of an overdensity region, which can become a favourable site for the 
formation of dust.

An alternative explanation for the large optical depths derived for 
these stars is that their distances are underestimated. Thus, their
luminosities would be significantly higher than the best values reported in
K22. The Q2 flag for all four sources may justify this consideration, as it makes indicates that their distances are highly uncertain.

We propose that the luminosities of these sources
follow the same $\tau_{10}-{\rm L}$ pattern traced by the carbon stars
in the sample that did not experience any HBB. In this case, ID 3 and
ID 4 would share a similar origin, as they are the brightest carbon stars in the
LMC investigated in T22, with luminosities in the $12000$ to $15000~{\rm L}_{\odot}$
range. Therefore, these stars would have evolved from
$2.5-3~{\rm M}_{\odot}$ progenitors that experienced a series of TDU episodes that favoured significant surface
enrichment of $^{12}$C and the s-process, which is in agreement with the surface chemical
composition derived by the observations (K22).
For ID 4, this explanation finds further confirmation in
the derived location of the dusty region around the star, which would
be similar to what was found for the bright carbon stars in the LMC.
The situation for ID 3 is more cumbersome on this side because the
inner border of the dust layer would be 
($4.5 \times 10^5~{\rm R}_{\odot}$) away, which
is significantly higher than what was found for the LMC counterparts.

We propose a similar interpretation for ID 7, which has a
luminosity of $\sim 8000~{\rm L}_{\odot}$, consistent with a 
$\sim 2~{\rm M}_{\odot}$ progenitor. In this case, the
distance of the dusty region would be $2.8 \times 10^5~{\rm R}_{\odot}$,
as shown in the right panel of Fig.\ref{fsumm}. 

\section{Mass loss and dust formation during the late AGB phases of
low-mass carbon stars}
\label{id817}
The post-AGB stars discussed in Section \ref{lowmc} were identified as the
progeny of low-mass carbon stars.
In the study of the counterparts of these sources in the LMC, the authors of T22
concluded that the dust observed in the present epoch was released after the general 
contraction of the star began, when the effective temperature increased to $3500-4000$ K.
Their analysis also showed that the mass-loss rates at the TAGB
required to reproduce the currently observed IR excesses are a factor of three 
higher than those found on the basis of stellar evolution modelling.

We now discuss the application of the analysis of T22 to the sources ID 8 and ID 17.
These stars are characterised by having the lowest and the highest optical depths among the
sub-sample considered in Section \ref{lowmc}. Study of these stars revealed the evolution of low luminosity post-AGB carbon stars
with different IR excesses. Because these stars were not investigated in 
detail in K23, we first interpreted their evolution based
on the observed physical and chemical properties then analysed
the dust responsible for the IR excess currently observed.

\begin{figure*}
\begin{minipage}{0.48\textwidth}
\resizebox{1.\hsize}{!}{\includegraphics{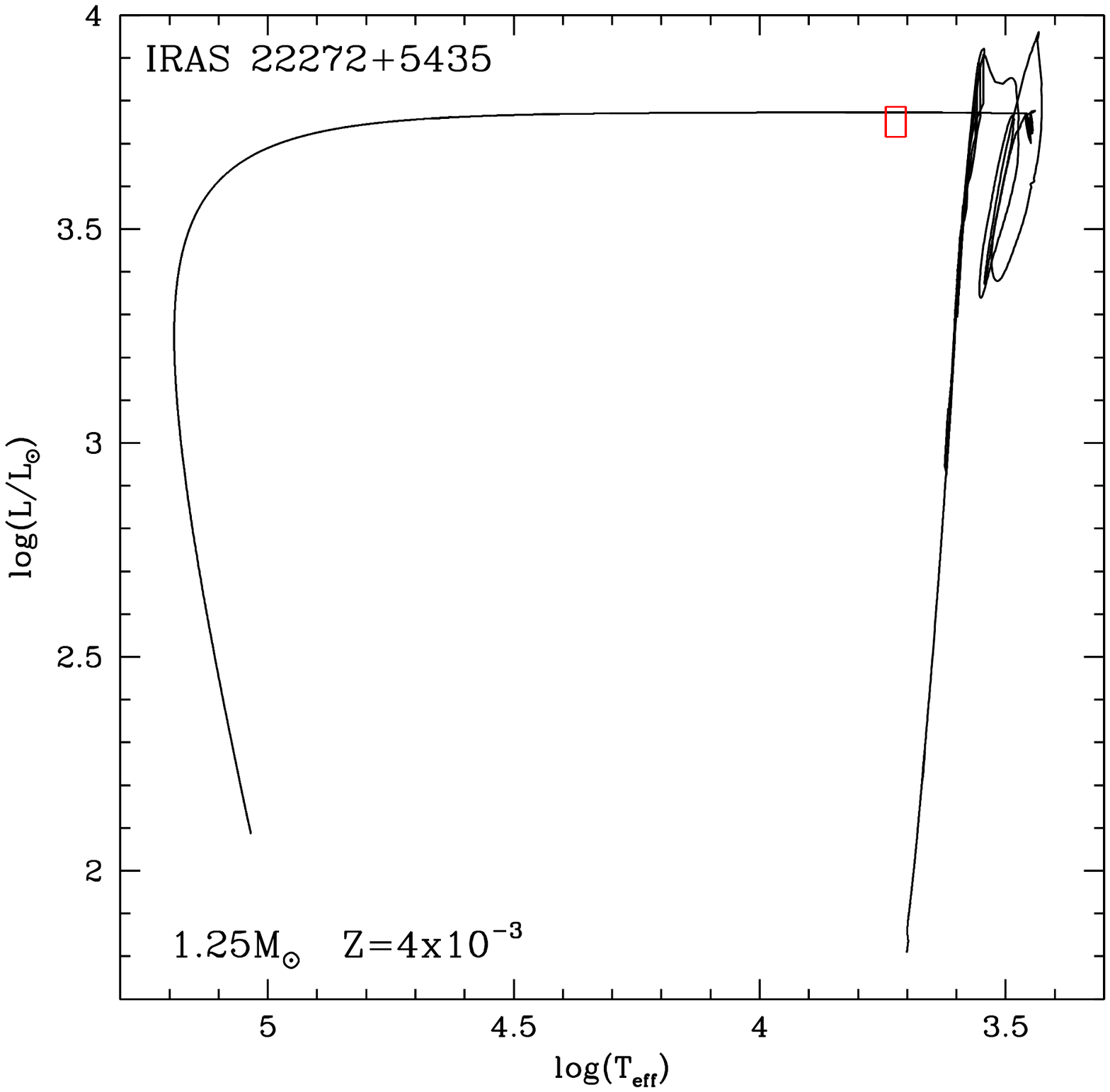}}
\end{minipage}
\begin{minipage}{0.48\textwidth}
\resizebox{1.\hsize}{!}{\includegraphics{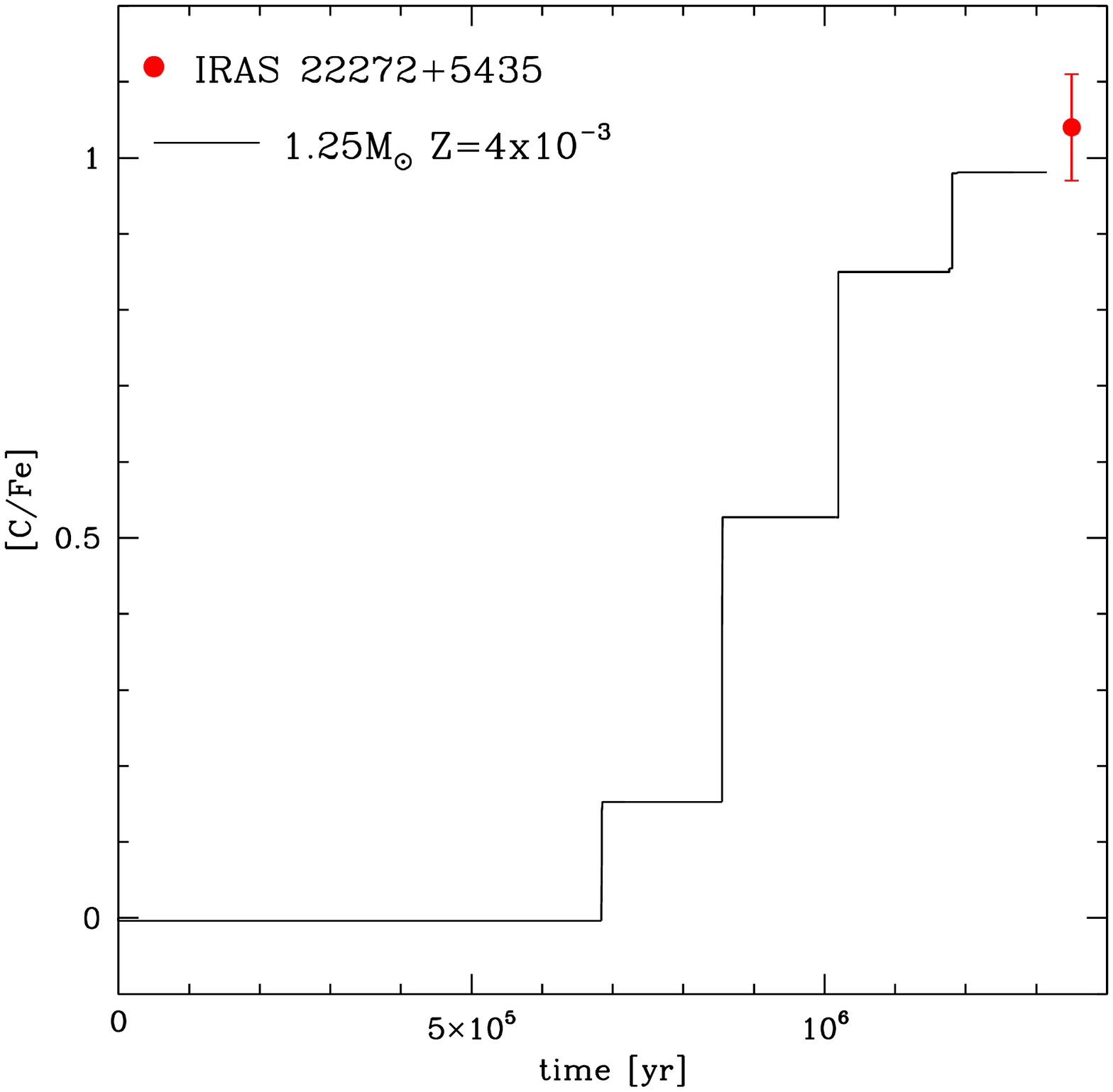}}
\end{minipage}
\vskip-60pt
\caption{Evolutionary track of a $1.25~{\rm M}_{\odot}$
star of Z $=4\times10^{-3}$ (left panel) and the variation of the surface
carbon of the same $1.25~{\rm M}_{\odot}$ model star during the AGB phase (right panel).
The red box in the left panel indicates the effective temperature and
luminosity (with the corresponding error bars) given for ID 17 by K22,
whereas the red data point in the right panel indicates the 
surface $[$C$/$Fe] of ID 17.}
\label{fid17}
\end{figure*}

\subsection{IRAS 22272 (ID 17): Past evolutionary history and 
the properties of the dust shell}
The left panel of Fig.~\ref{fid17} shows the evolutionary track of a
$1.25~{\rm M}_{\odot}$ model star with metallicity $Z=4\times 10^{-3}$
that evolved from the core helium burning, post-flash phase, to the white dwarf cooling sequence. 
The right panel of Fig.~\ref{fid17} shows how the surface carbon increased during two TDU events, which took place after the last two  experienced TPs.
Both the position of the star in the HR diagram and the 
derived surface carbon
are consistent with a progenitor of mass around $1.25~{\rm M}_{\odot}$, 
which agrees with the interpretation of
this source given in K23. This mass indicates that ID 17 formed 3-3.5 Gyr ago.

Regarding the dust in the surroundings of the star, Table 
\ref{tabero} shows that it is currently located at a distance
${\rm R}_{\rm in}=1.9 \times 10^5~{\rm R}_{\odot}$, and 
it is characterised by the optical depth 
$\tau_{10}^{\rm now}=8.7\times 10^{-3}$. The mass-loss rate at the
TAGB found via stellar evolution modelling is 
$\sim 1.5 \times 10^{-5}~{\rm M}_{\odot}/$yr. The models follow the standard 
description of mass loss by \citet{wachter08}, which was used to build
the evolutionary sequence shown in Fig.~\ref{fid17}. 
The authors of T22 found that the dust around low-mass post-AGB stars in the 
MC was released when the effective temperature 
was approximately 3500 K. Modelling the dust formation at the same
evolutionary phase of ID 17 resulted in an optical depth of $\tau_{10}^{\rm onset}=0.1$, which, after applying Eq.~1, corresponds
to a current value of $\tau_{10}^{\rm now}\sim 10^{-3}$.
This value is significantly smaller than the value deduced from 
the SED fitting ($8.7\times 10^{-3}$).

Consistency amongst the observations is found if we assume
that the mass-loss rate at the TAGB phase of ID 17
is a factor of approximately three higher than the value given above. This hypothesis leads to
$\tau_{10}^{\rm onset}=0.9$ when ${\rm T}_{\rm eff}=3500$ K, which, once more
by application of Eq.~1, corresponds to a current 
optical depth of $\tau_{10}^{\rm now}=8\times 10^{-3}$ 
and is consistent with the result from the SED fitting.
Under this assumption, the time required for the star to evolve
from the ${\rm T}_{\rm eff}=3500$ K until the present epoch
is approximately 300 yr, which is the time required
by the dust cloud to reach the current location if a 
10 km$/$s velocity is adopted.

\begin{figure*}
\begin{minipage}{0.48\textwidth}
\resizebox{1.\hsize}{!}{\includegraphics{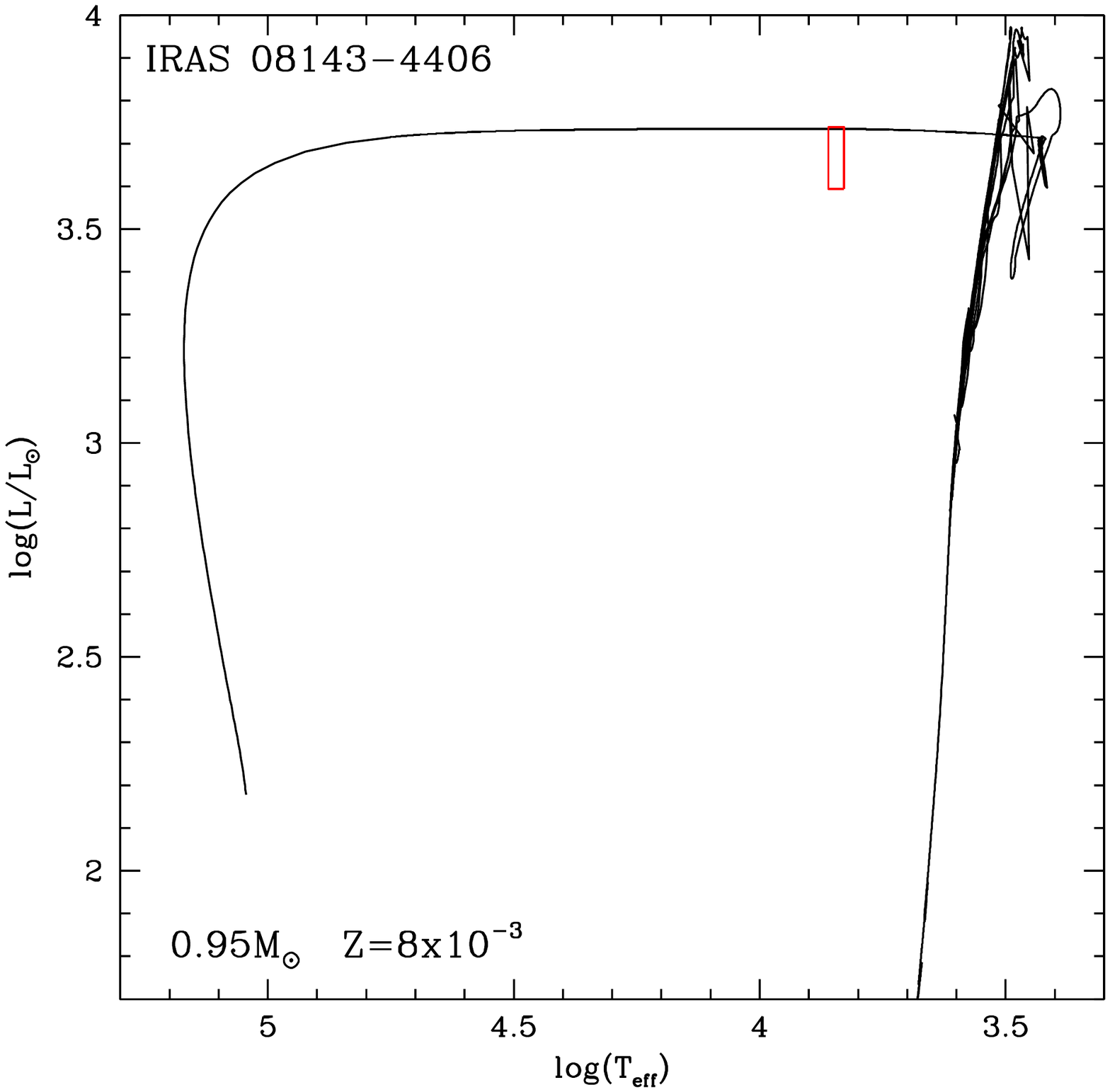}}
\end{minipage}
\begin{minipage}{0.48\textwidth}
\resizebox{1.\hsize}{!}{\includegraphics{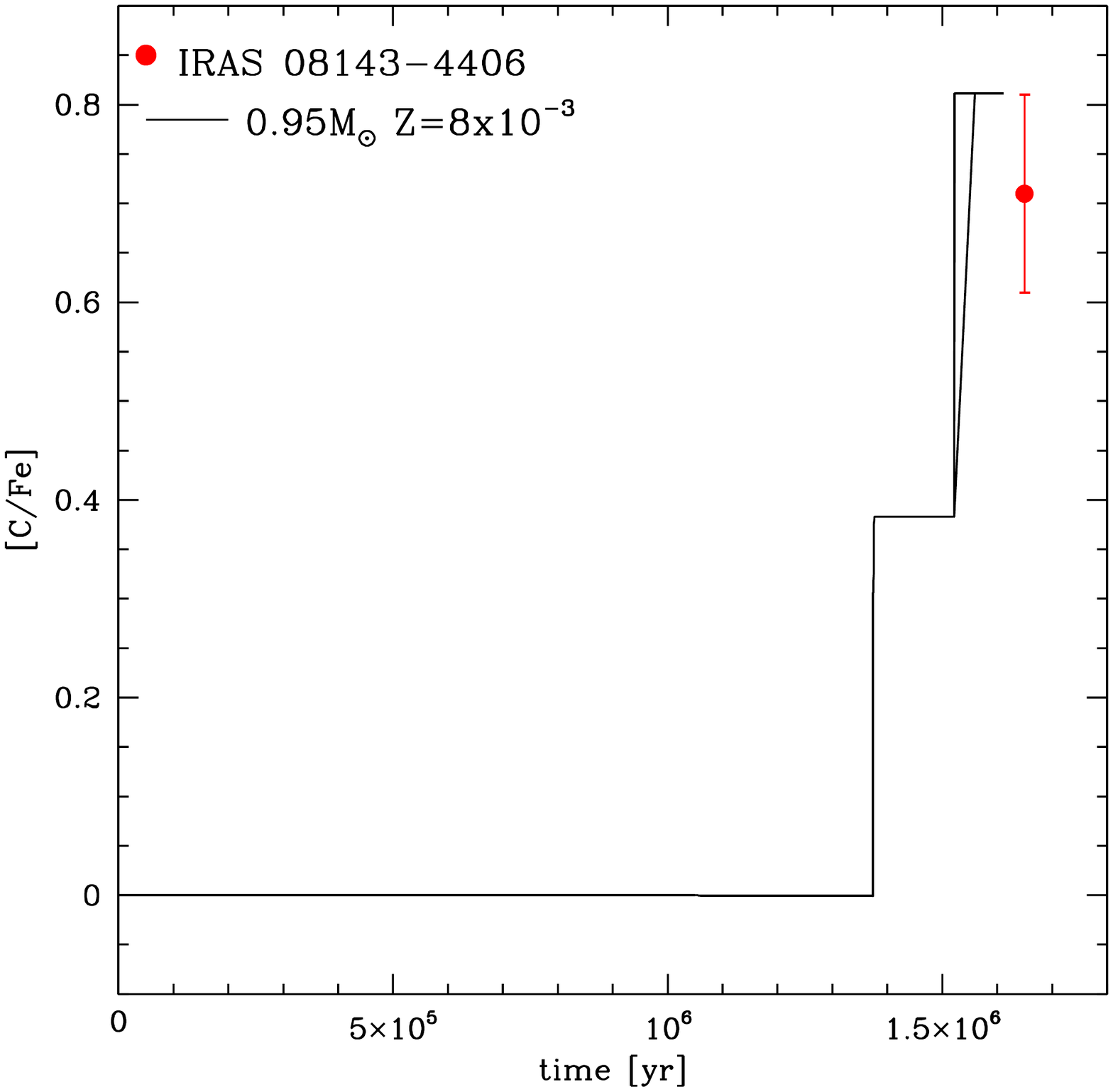}}
\end{minipage}
\vskip-60pt
\caption{Evolutionary track of a $0.95~{\rm M}_{\odot}$
star of $Z=8\times 10^{-3}$ (left panel) and the variation of the surface
carbon of the same $0.95~{\rm M}_{\odot}$ model star during the AGB phase (right panel).
The red box in the left panel indicates the effective temperature and
luminosity (with the corresponding error bars) given for ID 8 by K22,
whereas the red data point in the right panel indicates the 
surface $[$C$/$Fe] of ID 8.}
\label{fid8}
\end{figure*}

\subsection{The transition from the AGB to the post-AGB phase for 
IRAS 08143 (ID 8)}
Figure~\ref{fid8} shows that the position on the HR diagram and the surface 
carbon of ID 8
are consistent with the evolution of a 
$0.95~{\rm M}_{\odot}$ model star of metallicity $Z=8\times 10^{-3}$
and is in agreement with the interpretation given in K23. The observations and the results from stellar evolution modelling agree satisfactorily, which suggests that this source formed  approximately 11-12 Gyr ago. The final surface carbon
results from the two TDU events that happened after
the last two TPs, of which there were a total of eight.

The SED fitting for this source gave a current optical depth of $\tau_{10}^{\rm now}=2\times 10^{-3}$ and a distance of the dusty
region from the star ${\rm R}_{\rm in}=6.8 \times 10^5~{\rm R}_{\odot}$,
as reported in Table \ref{tabero}.
The conditions at the TAGB derived from stellar evolution modelling
are the effective temperature ${\rm T}_{\rm eff}=2650$ K, the stellar
radius $340~{\rm R}_{\odot}$, and the mass-loss rate 
$\dot{\rm M}^{\rm TAGB}=1.8\times 10^{-5}~{\rm M}_{\odot}/$yr.

When the scaling of $\dot{\rm M}$ with ${\rm T}_{\rm eff}$ proposed
in T22 is adopted, the crossing time from the TAGB to the
current evolutionary stage at ${\rm T}_{\rm eff}=7000$ K is 4000 yr. 
To cover the ${\rm R}_{\rm in}$ distance
found via the SED fitting given above, this would require
outflow velocities of the order of 3 km$/$s. It is more likely that the dust currently observed formed when the star began contracting and the radius 
decreased to $270~{\rm R}_{\odot}$, with a corresponding effective temperature
of $3000$ K. In this case, the crossing time would be
1500 yr, thus implying more realistic velocities of the wind on
the order of 10 km/s. 

Modelling the dust formation at the physical conditions corresponding to ${\rm T}_{\rm eff}=3000$ K leads to an optical depth of $\tau_{10}^{\rm onset}=0.4$. This starting point yields a current optical depth of $\tau_{10}^{\rm now}=1.8\times 10^{-3}$, which we consider to be in satisfactory agreement with the results from the SED fitting.

These results are based on
the mass-loss rate at the TAGB found via standard evolution modelling.
This approach differs from the analysis done earlier in this section
for ID 17 and, more generally, from the findings in T22 regarding 
low-mass carbon stars, in which a factor of three increase in 
$\dot{\rm M}^{\rm TAGB}$ was required to simulate the results
obtained from the SED fitting.

We conclude that the mass-loss rates derived for low luminosity
carbon stars on the basis of the treatment by \citet{wachter08}
can be safely used for solar or slightly sub-solar chemistry when
the carbon excess with respect to oxygen is small. On the
other hand, this approach underestimates the mass-loss rates of the metal-poor
counterparts of the same luminosity.

\section{Conclusions}
\label{concl}

We modelled the stellar evolution and dust formation of a sample of carbon-rich, post-AGB single stars in the Galaxy (see K22). We characterised individual sources with radiative transfer modelling, derived their age and progenitor mass, and reconstructed their past evolutionary history starting from the final AGB phases.

Radiative transfer modelling probes the dust mineralogy around the individual sources, the optical depth, and the distance of the dusty region from the 
central star. This information was used to reconstruct the past evolutionary history from the late AGB phase to the present time so that we could deduce the timescale of the AGB to post-AGB transition and study the efficiency of the dust formation process during the final evolutionary phases. Our analysis has revealed new insights into the evolution of
stars on and away from the AGB in the Galaxy.

Most of the sources investigated evolved from low-mass 
(${\rm M}<1.5{\rm M}_{\odot}$) stars that became carbon stars after
a series of TDU events. The optical depth of these sources is generally anti-correlated with metallicity. 
This is because metal-poor stars reach higher carbon-to-oxygen excesses
than their more metal-rich counterparts of similar luminosity, due to the lower amount of oxygen initially locked in the star. This study confirms the results from previous investigations focused on carbon-rich post-AGB stars in the Magellanic Clouds, where the optical depth is correlated with luminosity. For Magellanic objects, the evolution to the post-AGB phase is faster for more luminous stars, which moves the dusty region closer to the central star in the brightest sources. 

Two stars in our sample evolved from $3-4~{\rm M}_{\odot}$ progenitors that
experienced both TDU and HBB. Their IR excess is smaller than observed in the 
slightly lower-mass counterparts that experienced TDU only because HBB reduces the final carbon excess with respect to oxygen.

Our sample also includes four stars that, on the basis of SED fitting and of the 
recommended $\it{Gaia}$ DR3 parallaxes, have luminosities significantly 
lower than the threshold required to become carbon-rich stars. These stars are also characterised 
by a large IR excess. We propose that the uncertain parallaxes lead to underestimated distances and that these sources are indeed the progeny of $2-3~{\rm M}_{\odot}$ 
stars that became carbon rich after several TDU episodes.

 The present study provides important information on the mass loss experienced
by carbon-rich stars at the end of the AGB evolution. To reproduce the currently observed 
IR excesses, the mass-loss rates during the final AGB phases must be of the order of 
$4-5\times 10^{-5}~{\rm M}_{\odot}/$yr, which is approximately three times higher than what has been found by standard 
evolution modelling. However, the mass-loss rates expected for solar 
chemistry low-mass stars evolving through the final AGB phases of the order of 
$1-1.5\times 10^{-5}~{\rm M}_{\odot}/$yr are consistent with the observational scenario. 
These findings confirm the results obtained by the authors of T22 and indicate the need for an
improved treatment of the mass-loss mechanism by carbon stars, which must take into account 
not only the physical parameters of the star during a given evolutionary phase but also
the surface chemical composition.


\begin{acknowledgements}
ST acknowledges the anonymous referee for the very constructive and helpful comments and suggestions.  
DK acknowledges  the  support  of  the  Australian  Research Council (ARC)  Discovery  
Early  Career  Research  Award (DECRA) grant (DE190100813). This research was supported 
in part by the Australian Research Council Centre of Excellence for All Sky Astrophysics 
in 3 Dimensions (ASTRO 3D), through project number CE170100013. FDA and PV acknowledge the 
support received from the PRIN INAF 2019 grant ObFu 1.05.01.85.14 (“Building up the halo: 
chemo-dynamical tagging in the age of large surveys”, PI. S. Lucatello). 
HVW acknowledges support from the Research Council of the KU Leuven under grant number C14/17/082. 
EM acknowledges support from the INAF research project “LBT - Supporto Arizona Italia".

\end{acknowledgements}

\appendix
\section{MgS dust}
We tested the possibility that MgS dust is responsible for the formation of the
feature centred at $30~\mu$m in the SWS spectra of the investigated post-AGB sample.
To this aim, we built additional synthetic SEDs for the source ID 14 (selected as it is among those
with the most prominent $30~\mu$m feature) in which approximately 10\% of MgS dust was included. The 
results of such an exploration can be seen in the top-central panel of Fig.~\ref{fsed2}. 

We find that when pure MgS dust is considered, the corresponding feature is too sharp (see green line in 
Fig.~\ref{fsed2}), with little or no improvement with respect to the case with no MgS (red line). 
We further considered the possibility proposed by \citet{mgs}, and extensively applied by \citet{ester21} to study the C-star AGB population of the LMC, that MgS dust grows around SiC cores. In this case, we obtained the
synthetic SED indicated with a blue line in Fig.~\ref{fsed2}, which shows a much better
consistency with the SWS spectra. For the sake of completeness, we report that
the aforementioned result was obtained by assuming that the MgS mantle has the
same width as the SiC core, around 0.07 $\mu$m. This is not fully justified on
the basis of the thermodynamic properties of the MgS dust \citep{chase98}, which is expected
to form in an external zone of the circumstellar envelope of carbon stars
\citep{mgs};
thus the growth of the SiC-MgS grains should be limited to approximately 30\% of
the SiC core. Yet we believe this numerical experiment shows the possibility of MgS dust being the main actor in the formation of the
$30\mu$m is fully reasonable, although a deeper exploration is required
before this hypothesis can be confirmed. We leave this problem open.

%
%

\end{document}